\documentclass[useAMS,usenatbib,a4paper]{mn2e}
\usepackage{amssymb,amsmath}
\usepackage{graphicx}
\usepackage{natbib}
\usepackage{journal_shortcuts}

\begin{document}
\def\crp{CRp }
\def\cre{CRe }
\def\eps{\varepsilon}
\def\e{{\rm e}}
\def\p{{\rm p}}

\title{Magnetic field amplification by cosmic-ray-driven turbulence: I. isotropic CR diffusion}
\author[M. Br\"uggen]{M. Br\"uggen$^{1}$\thanks{E-mail: mbrueggen@hs.uni-hamburg.de}\\ \\
$^{1}$Hamburger Sternwarte, Universit\"at Hamburg, Gojenbergsweg 112, 21029 Hamburg, Germany}
\maketitle

\label{firstpage}

\begin{abstract}
We have performed magnetohydrodynamical simulations to study the amplification of magnetic fields in the precursors of shock waves. Strong magnetic fields are required in the precursors of the strong shocks that occur in supernova remnants. Observations also suggest that magnetic field amplification takes place in the weak shocks that occur in galaxy clusters and that produce so-called radio relics. Here, we extend the study of magnetic field amplification by cosmic-ray driven turbulence to weak shocks. The amplification is driven by turbulence that is produced by the cosmic-ray pressure acting on the density inhomogeneities in the upstream fluid. The clumping that has been inferred from X-ray data for the outskirts of galaxy clusters could provide some of the seed inhomogeneities. Magnetic field power spectra and Faraday maps are produced. Furthermore, we investigate how the synchrotron emission in the shock precursor can be used to verify the existence of this instability and constrain essential plasma parameters.
\end{abstract}


\section{Introduction}

Radio relics are diffuse synchrotron sources in galaxy clusters. They are polarised (generally at a level of $10$ to $30\%$) and typically show a steep spectrum. Moreover, they have elongated shapes, with sizes between $100$ kpc and $\sim 2$ Mpc, and are usually located at the cluster periphery.
Relics are thought to be caused by shocks, triggered by a major merger and propagating through the intra-cluster medium (ICM). In a few cases (A521, \citealt{gvm08}; A3667, \citealt{fsn10}; A754, \citealt{mmg11}) the radio emission in relics is co-located with a shock front, detected by X-ray observations.
Recent observations have given  support to the hypothesis that they trace shock fronts \citep{vanweeren10}, even though some puzzles remain \citep{ogrean13}. At shock fronts, particles can be accelerated, e.g., via the diffusive shock acceleration mechanism \citep[DSA;][]{1977DoSSR.234R1306K, 1977ICRC...11..132A, 1978MNRAS.182..147B, 1978ApJ...221L..29B}. The resulting steady-state CR distribution is a power-law in momentum, $p$, ($f \propto p^{-\alpha}$) with a slope $\alpha = 4M^2/(M^2-1)$, where $M$ is the shock flow Mach number. 
Non-linear diffusive shock acceleration is subject to intense research using high-performance computers and some details of it are currently under revision (see \citealt{caprioli10} for a recent review).

Observations suggest that radio relics host relatively large magnetic fields, with typical values at the $\mu$G level. These fields are derived using several methods, mostly equipartition arguments, comparison of hard X-ray and radio emission and spectral modelling \citep{markevitch05, vanweeren11}. Magnetic field strengths of this magnitude are known to exist in the central regions of clusters but are not expected for the peripheral regions in which radio relics are found. An important process for magnetic field amplification, often invoked in the framework of the evolution of galaxy clusters, is the small-scale, turbulent dynamo \citep{dolag02,bruggen05,subramanian06}. Recently, \cite{iapichino12} discussed various sources for magnetic fields in radio relics. Ignoring cosmic-ray driven effects, it was concluded that turbulence produced by the shock itself cannot explain the magnitude of these fields. However, it was argued that if the turbulent pressure support in the ICM upstream of the merger shock is of the order of 10-30 per cent of the total pressure on scales of a few times 100 kpc, then vorticity generated by compressive and baroclinic effects across the shock discontinuity could lead to a sufficient amplification of the magnetic field. For a typical shock Mach number $M \sim 3$, the resulting amplification factor due to compressional amplification was found to be  $\sim 2.5$. If the upstream magnetic fields are below $\sim \mu$G this amplification is insufficient to explain the fields found in radio relics.

Diffusive shock, or Fermi-I, acceleration is also presumed to be efficient in supernova remnants (SNRs). However, in SNRs upstream magnetic fields of around 5 $\mu$G, typical of the interstellar medium (ISM), are too weak to provide an efficient acceleration of the cosmic rays with energies as high as $10^{15}$ GeV, as required for supernova shocks to produce CRs to the knee. PeV cosmic rays have long mean free paths in such a field and have a high probability of escaping the shock, so they are not subject to further acceleration. In SNRs the magnetic fields in the shocked gas is much higher than typical fields found in the interstellar medium (e.g. \citealt{vink03}). 

Current-driven instabilities  \citep{bell13, niemiec08, bell04} have been proposed to amplify magnetic fields by large factors. The electric current that drives this instability comes from the drift (streaming) of the escaping CRs. The compensating return current of the background plasma leads to a transverse force on the background plasma that can amplify transverse perturbations in the magnetic field. This mechanism has the draw-back that it will only generate magnetic fields on spatial scales that are smaller than the gyro-radius of the streaming particles.  Without, for example, an inverse turbulent cascade these fields have scales that are too small to participate in the acceleration of the highest energy particles. A number of numerical simulations have shown that the magnetic field can be amplified by a factor of $\sim 10 - 45$ via  Bell's CR current-driven instability (Riquelme \& Spitkovsky 2009, 2010; Ohira et al. 2009). In the test-particle regime where the flow structure is not modified, it is found that the amplification of the magnetic field is proportional to the Alfvenic Mach number for the far upstream Alfven speed.  Moreover, it was suggested that long-wavelength magnetic fluctuations can grow also in the presence of short-scale, circularly-polarized Alfven waves excited by the Bell-type instability \citep{bykov11}.\\

Inoue (2009) have simulated the interaction of a strong shock with a multi-phase medium and studied the magnetic field amplification in the postshock region. 
\citet{beresnyak09} proposed an analytic model, in which the cosmic-ray pressure drives turbulence which, in turn, amplifies the pre-shock magnetic field.  

The turbulence arises from the fact that the cosmic ray pressure gradient in the shock precursor exerts a force on the upstream plasma that is in general not proportional to the gas density. 
Density fluctuations then lead to acceleration fluctuations, which then induce further density fluctuations.  In one dimension, the instability vanishes if the diffusion coefficient for the cosmic-rays happens to be inversely proportional to the mass density. However, it is impossible to suppress the instability in more than one dimension.  In case that the distribution of the cosmic-ray pressure is chosen to avoid the instability perpendicular to the shock front, it is unstable parallel to the shock front and vice-versa.  The effective diffusion coefficient caused by the scattering experienced by the cosmic rays,   is a complex and poorly understood function of the local magnetic field strength the power-spectrum of magnetic irregularities.  Consequently, the diffusion coefficient will change if the plasma is locally adiabatically compressed, and this will, in turn, alter the cosmic-ray pressure gradients. This small-scale dynamo saturates only by non-linear effects when magnetic field pressure approaches equipartition with the kinetic energy density. Finally, \citet{beresnyak09} find that solenoidal forcing is the most efficient mode for turbulent amplification, which is in agreement with results by \cite{federrath11}.\\

\cite{drury12} present two-dimensional numerical simulations which demonstrate that magnetic field amplification via CRs by factors of 20 or more can be achieved under conditions typical of the ISM and supernova blastwaves. The length scale over which this amplification occurs is that of the diffusion length of the highest energy non-thermal particles.

While the ISM is known to have a multi-phase composition with significant density inhomogeneities, it is less clear what can produce the seeding in the ICM.
\cite{simionescu11} determined ICM properties out to the edge of the Perseus Cluster. The apparent baryon fraction exceeds the cosmic mean at larger radii, suggesting a clumpy distribution of the gas for radii larger than half of the virial radius.
In general, gas clumping may constitute a source of uncertainty in the derivation of properties of galaxy clusters atmospheres since a significant part of detected photons
may come from the most clumpy structures of the ICM, which may not be fully representative of the underlying large-scale distribution of gas \citep[e.g.][]{eckert12}.  
Indeed, a significant fraction of the gas mass in cluster outskirts may be in the form of dense gas clumps, as suggested by recent simulations \citep[e.g.][]{nagai11,vazza13}. \\

Here, we simulate the amplification of magnetic fields by interaction of cosmic-rays and density inhomogeneities. We perform 2D and 3D simulations of weak and strong shocks, at which we inject cosmic rays that are subject to advection and diffusion. The ones that diffuse ahead of the shock interact with the density inhomogeneities and can amplify the pre-shock magnetic field. Clearly, the diffusion of cosmic rays is governed by the topology of the magnetic field and will in general be anisotropic. The diffusion coefficients in conditions as found in the ICM are highly uncertain and subject to current research \citep[e.g.][]{casse02}. Hence, in this first paper of this series, we will assume isotropic diffusion of CRs.\\

In Sec. 2, we describe the code and the simulation setup. In Sec.~3 we present our results and we conclude in Sec.~4.

\section{Method}

We performed our simulations using FLASH4.0, a parallel 
magnetohydrodynamics/N-body astrophysical simulation code developed at the Center for Astrophysical Thermonuclear Flashes 
at the University of Chicago (Fryxell et al. 2000). FLASH uses adaptive mesh refinement (AMR), a technique that places 
higher-resolution elements of the grid only where they are  needed.  Magnetic fields were computed using the unsplit staggered mesh method \citep{lee09} which is based on a finite-volume, high-order Godunov scheme (3rd order PPM) with 
constrained transport. A major strength of the unsplit-mesh algorithm is that it preserves vanishing divergence of the magnetic field
to very high accuracy. For simplicity we neglected cooling processes of the gas, such as and the cosmological expansion of space, both of which will have a negligible effect in within the time scales studied here. The gas is assumed to be ideal with a polytropic equation of state with a ratio of specific heats of 5/3.

In conservation form, the equations that are solved are
\begin{eqnarray}
\frac{\partial \rho}{\partial t} + \mathbf{\nabla} \cdot \left(\rho
		      \mathbf{u}\right)  & = & 0 \label{mass} \\
\frac{\partial \rho \mathbf{u}}{\partial t} + \nabla\cdot\left( \rho \mathbf{u} 
	\mathbf{u} + P\right) + \frac{4\pi}{c}\mathbf{B}\times(\nabla\times\mathbf{B}) & = & 
\mathbf{F}_{\rm cr}, \label{neutral_mom} \\
\frac{\partial e}{\partial t} + \mathbf{\nabla} \cdot \left[ \left(e +
		P+\frac{B^2}{2}\right)\mathbf{u}-\mathbf{B}(\mathbf{v}\cdot\mathbf{B})\right] & = &
\mathbf{F}_{\rm cr}\cdot\mathbf{u}\\
\frac{\partial \mathbf{B}}{\partial t} - \mathbf{\nabla}\times \mathbf{u}\times \mathbf{B} & = & 0, \label{B_eqn} \\
								  \nabla\cdot\mathbf{B} & = & 0 \label{divB}
\end{eqnarray}
where $\rho$ is the mass density, $e=p/(\gamma - 1)+\rho u^2/2 +B^2/2$ the energy density, $\mathbf{u}$ is the fluid velocity,
$P$ is the thermal pressure, $\mathbf{B}$ is the magnetic field and
$\mathbf{F}_{\rm cr}$ is the force per volume due to the cosmic ray pressure
gradient.  $\mathbf{F}_{\rm cr}$
is given by
\begin{equation}
\mathbf{F}_{\rm cr}  =  - \mathbf{\nabla} P_{\rm C} . 
\end{equation}
We chose a computational box of physical size 2 $\times$ 1 $\times$ 1 in non-dimensional units. In 2D we used a grid of dimension 2 $\times$ 1.  For our fiducial simulations, we used 2 $\times$ 1 base blocks and seven levels of refinement, yielding an effective resolution of 1024 $\times$ 512 cells. The adaptive mesh refinement was triggered on standard criteria based on the gradients of gas density and the cosmic rays density. On the $x$-boundaries we chose inflow/outflow boundary conditions, and on all other boundaries we enforced reflecting boundaries.

The cosmic ray pressure is subject to advection and diffusion, i.e.:
\begin{equation}
\frac{\partial P_{\rm C}}{\partial t} + \mathbf{\nabla} \cdot \left(P_{\rm C}\mathbf{u}\right)  =  \mathbf{\nabla} \cdot (D\mathbf{\nabla}P_{\rm C}) .
\end{equation}
Motivated by \citep{drury12}, cosmic rays were injected by enforcing at all times that at the location of the shock $P_{\rm C}=\eta \rho_{\rm up} u_{\rm up}^2$. This is a value constant in time irrespective of how the precursor changes. Practically, the value $P_{\rm C}$ is fixed by adding to the cells that mark the shock front such that the cosmic-ray pressure retains the desired value. Somewhat more conservatively than Drury, we picked $\eta=0.1$.  At the start of the simulation, the CR pressure is zero everywhere, $P_{\rm C}=0$, and the shock position is at the centre of the box at $x=1$. 
 The boundary conditions for the cosmic ray fluid were outflowing on all sides. Thus, the efficiency, i.e. the fraction of the upstream ram pressure that was converted into cosmic rays is assumed to be a fixed fraction independent of the Mach number. 


A number of papers point to a strong dependence of acceleration efficiency on the shock Mach number, whereby stronger shocks produce higher efficiencies. Kang et al. (2007) report an investigation of acceleration efficiency behavior in diffusive shock acceleration simulations in quasi-parallel shocks with a Bohm diffusion coefficient, a self-consistent treatments of particle injection from the thermal pool into the acceleration process, and Alfven wave propagation.

Unlike in \cite{drury12}, we make no assumptions about the cosmic-ray pressure profile in the precursor but compute the pressure profile self-consistently by solving for the diffusion of the CRs from the fixed injection point at the original shock location.  
The diffusion equation for the cosmic ray fluid was solved using an unsplit general purpose diffusion solver that relies on the HYPRE library. It is advanced in time implicitly using a Crank-Nicholson scheme. Using units of  length of $D/u_{\rm up}$, we can scale our results for a range of values of $D$.  As customary and used e.g. in Kang \& Jones (2005), we normalize our spatial and temporal coordinates by $t=D/u_{\rm up}^2$ and $x=D/u_{\rm up}$, respectively. This approach removes the need for any particular choice in $D$. In the simulations described below we express pressures and energy densities in units $P=\rho u_{\rm up}^2$. For the example of cluster outskirts, we use as the unit of length to 100 kpc, unit of velocity $10^8$ cm s$^{-1}$ and unit of density $10^{-26}$ g cm$^{-3}$, which yields a unit of pressure of $10^{-10}$ erg cm$^{-1}$ s$^{-2}$ and unit of time of 3.08 $\times 10^{15}$ s or 98 Myr. The units for the wave number $k$ is also given in code units, i.e. a wavenumber of 1 corresponds to 1/length of box. For the field strengths considered here, $D$ could be given by the Bohm diffusion coefficient $D = m_p c^3/(3eB_0) = (3.13\times 10^{28}{\rm cm}^2{\rm s}^{-1})(B/\mu{\rm G})^{-1}$, which is within the range inferred for CRs in the ICM from the Coma radio halo \citep{schlickeiser87}. (see also table 3 in \citealt{bonafede11}). The form of the shock is assumed to be planar and fixed in space. In future work we are going to implement shock finding techniques to compute the shock surface at run-time.

Clearly, the diffusion coefficient as well as the injection rate depend strongly on the shock obliquity and diminishes as the angle between the ambient field and the shock normal increases (e.g. \cite{voelk03}). 
The perpendicular acceleration time at high energies is significantly smaller than the Bohm acceleration timescale, which implies expedient acceleration at perpendicular shocks.
Our fiducial simulations are performed with perpendicular shocks, but we also show some results for simulations with parallel shocks. \cite{zank06} construct a model for diffusive particle acceleration at highly perpendicular shocks, i.e., shocks whose upstream magnetic field is almost orthogonal to the shock normal. For the case of the solar wind, they find that parallel and perpendicular shocks can inject protons with equal facility. It is only at highly perpendicular shocks that very high injection energies are necessary. While particle acceleration at quasi-parallel shocks is reasonably well understood, particle acceleration at perpendicular shocks is largely unexplained. In particular, theoretical models for the diffusion coefficient perpendicular to the field lines do not appear to describe observations, nor are they consistent with careful numerical simulations \citep{giacalone06, qin02}.

As in \cite{drury12}, the initial distribution of density is defined via a function $f$ that is chosen to be
\begin{equation}
f(x,y,z) = \sum_{k_x,k_y,k_z}
A_i \sin(k_x x+\phi_i)\sin(k_y y+\xi_i)\sin(k_z z +\psi_i)
\end{equation}
where $i$ is an index for the wave-vector, $\mathbf{k}$. The amplitude $A_i$ and $\phi_i$, $\xi_i$, $\psi_i$ are
random variables picked from the ranges $[0,1]$ and $[0, 2 \pi]$, respectively.  The wave numbers, $k_x$, $k_y$ and
$k_z$ range from 10 to 50.  Then, the density 
distribution is given by

\begin{equation}
\rho(x,y,z) = e^{\beta f(x,y,z)} ,
\label{eqn:rho-def}
\end{equation}
where $\beta$ is a constant that is chosen to yield the desired RMS of the density distribution.
For all simulations shown here, the RMS is chosen to be 
0.2. This setup yields a log-normal distribution for the density distribution which is typical of what 
would be expected in the case of isothermal turbulence.

Assuming that the incoming flow contains density irregularities of magnitude $\delta \rho$ on a length scale $\lambda$ the cosmic-ray pressure, which operates on a time scale that is similar to the advection time through the precursor, will lead to velocity fluctuations of the order
\begin{equation}
\delta u \approx {\delta \rho\over \rho_0} {F_{\rm cr}\over\rho_0}{D\over u^2_{\rm up}}
\label{eqn:delta-u}
\end{equation}
on the same spatial scale $\lambda$ \citep{drury12}. Here the subscript zero denotes unperturbed quantities.  In order for turbulence to develop, the eddy turn-over time has to be short compared to the outer length-scale, which implies
\begin{equation}
{\lambda\over\delta u} \ll {D\over u^2_{\rm up}}   .
\label{eqn:lambda-limit}
\end{equation}

We ran simulations for three Mach numbers $M=2$ (runs with label M2),  $M=3$ (runs with label M3) and $M=100$ (runs with label 100).  These Mach numbers refer to the values of the initial conditions. The former two are representative of shocks in radio relics and the latter of shocks in supernova remnants.
The initial conditions were chosen to produce stationary shocks with left and right states $(\rho,p,u_x)_{\rm up}=(1,1,2.581988)$ and $(\rho,p,u_x)_{\rm down}=(2.2857143,4.75,1.1296198)$ for the $M=2$ shock, $(\rho,p,u_x)_{\rm up}=(1,0.0681818,1)$ and $(\rho,p,u_x)_{\rm down}=(3.,0.75,0.3333333)$ for the $M=3$ shock and $(\rho,p,u_x)_{\rm up}=(1,0.0000600,1)$ and $(\rho,p,u_x)_{\rm down}=(3.9988036,0.7499850,0.2500750)$ for the $M=100$ shock.
We only look at cases of weak, dynamically unimportant magnetic fields. The initial magnetic field chosen to be uniform and purely in the $y$ direction and so these simulations are appropriate for a perpendicular shock.

Note that for the initial conditions we use the ordinary Rankine-Hugoniot relations and not those modified by the presence of an additional cosmic-ray fluid and the magnetic field.

\begin{table}
\caption{Simulation parameters}
\begin{center}
\begin{tabular}{|c|c|}
Mach number &  eff. grid size  \\ \hline\\
100 & $1024\times 512$  \\
100 & $512\times 256$  \\
100 & $2048\times 1024$  \\
 100 &  $512\times 256 \times 256$ \\
3 & $1024\times 512$  \\
3 & $512\times 256 \times 256$  \\
2 & $1024\times 512$  \\
2 & $512\times 256 \times 256$  \\
\end{tabular}
\end{center}
\label{default}
\end{table}

\section{Results}

\subsection{Magnetic field amplification for different Mach numbers}

\begin{figure}
\begin{center}
\includegraphics[width=\columnwidth]{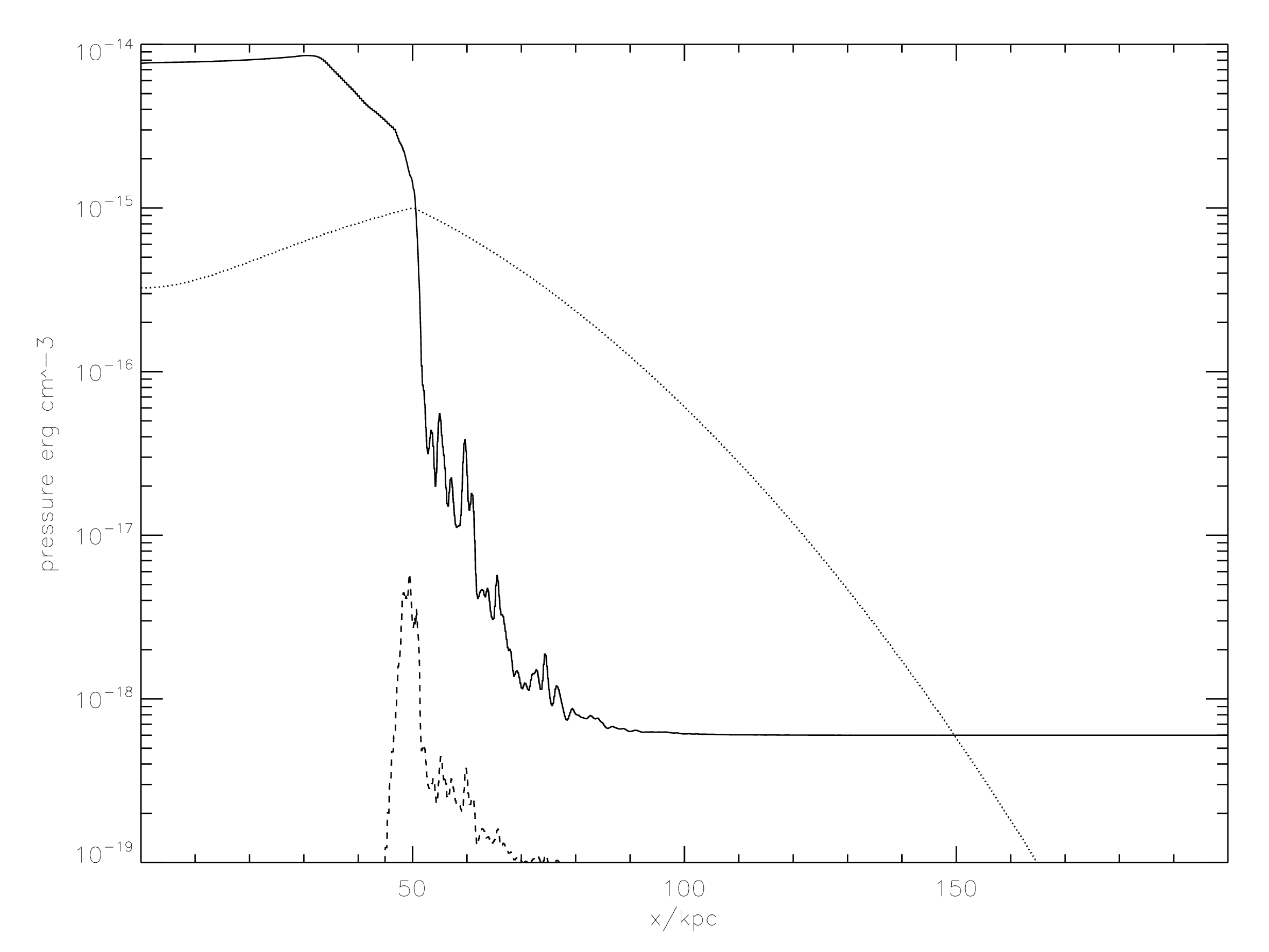}
\caption{Profiles of gas pressure (solid), cosmic-ray pressure (dotted) and magnetic pressure (dashed) for a $M=100$ shock in the vicinity of the shock transition. Units were scaled for conditions typical of the ICM.}\label{fig:pprof}
\end{center}
\end{figure}

\begin{figure*}
\begin{center}
\includegraphics[width=0.35\textwidth]{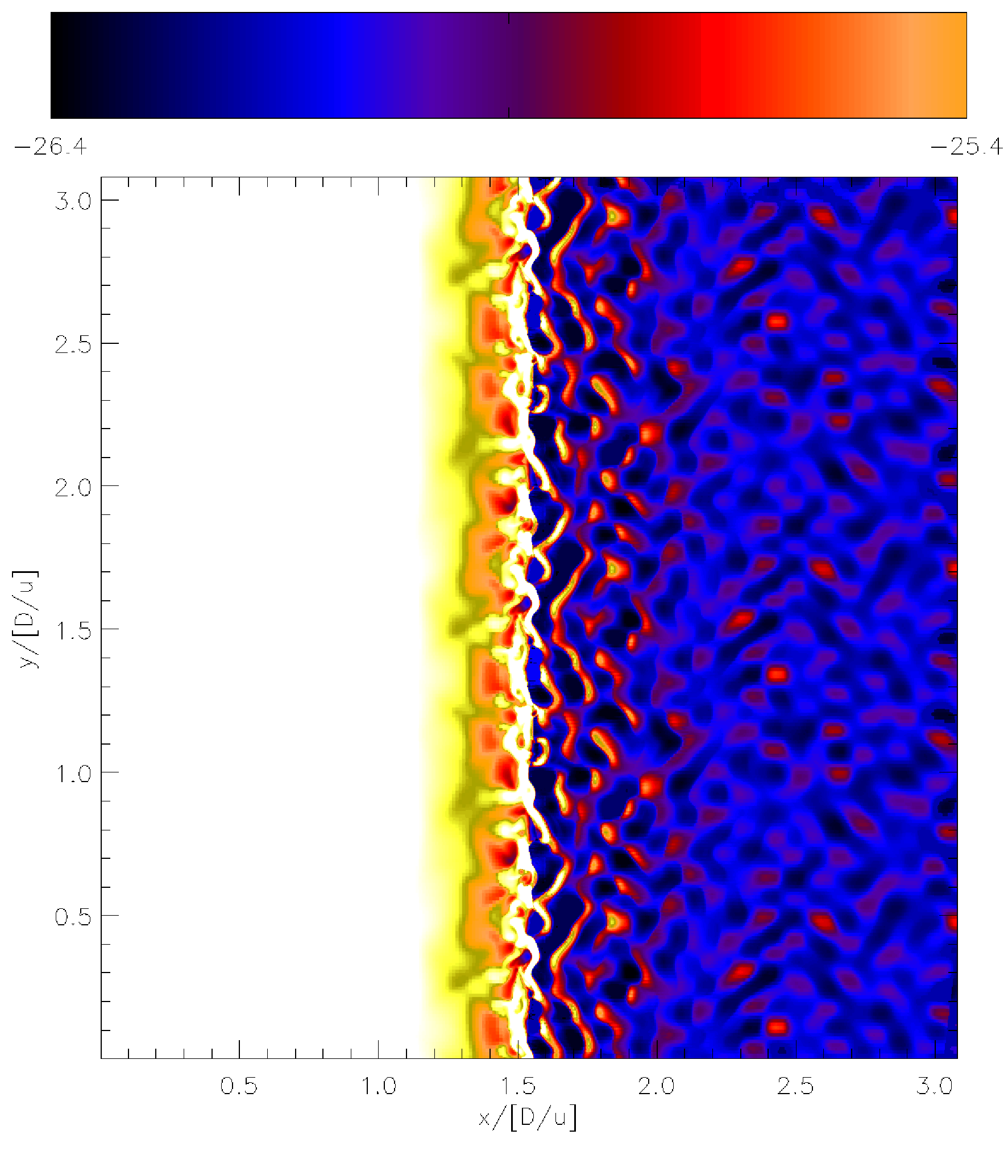}
\includegraphics[width=0.35\textwidth]{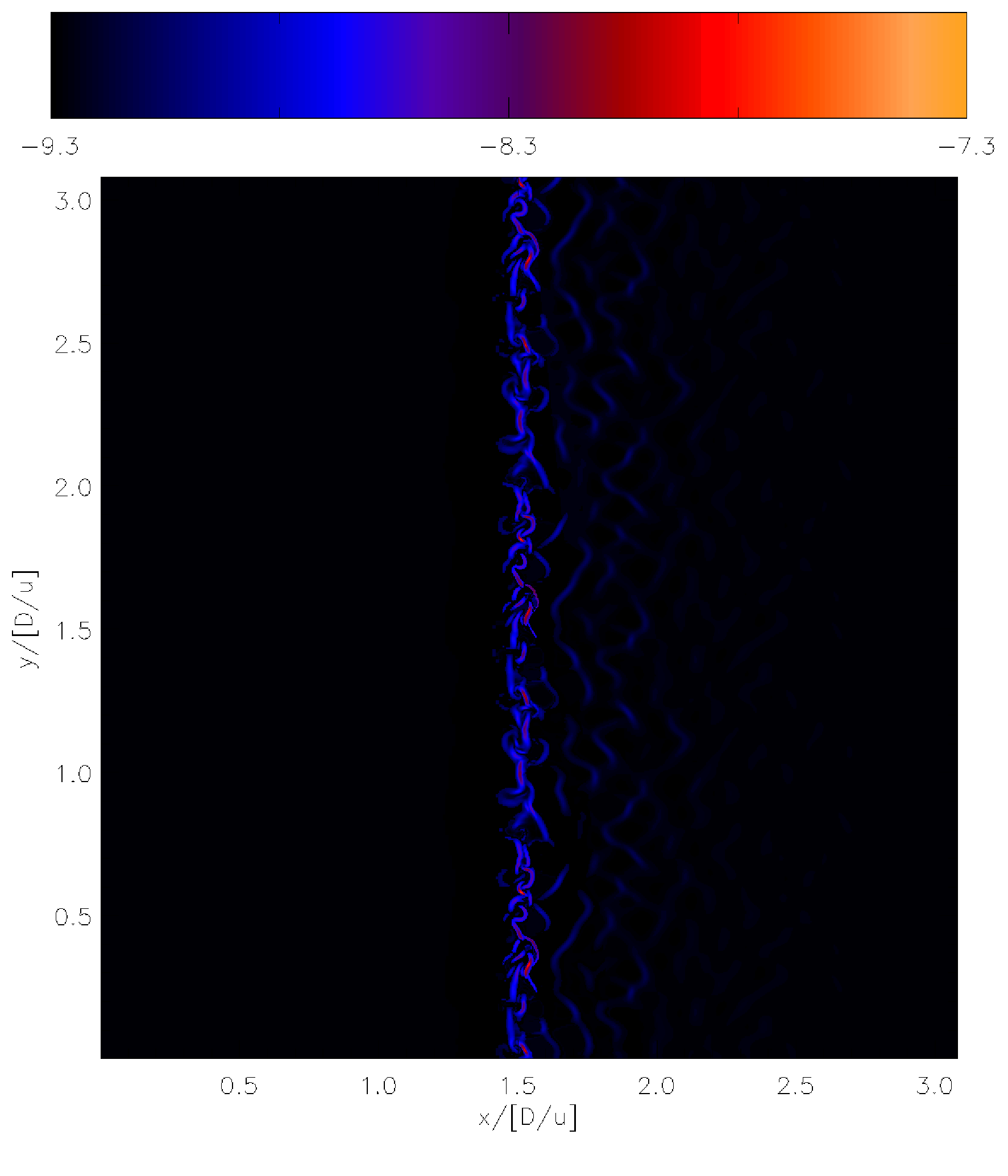}\\
\includegraphics[width=0.35\textwidth]{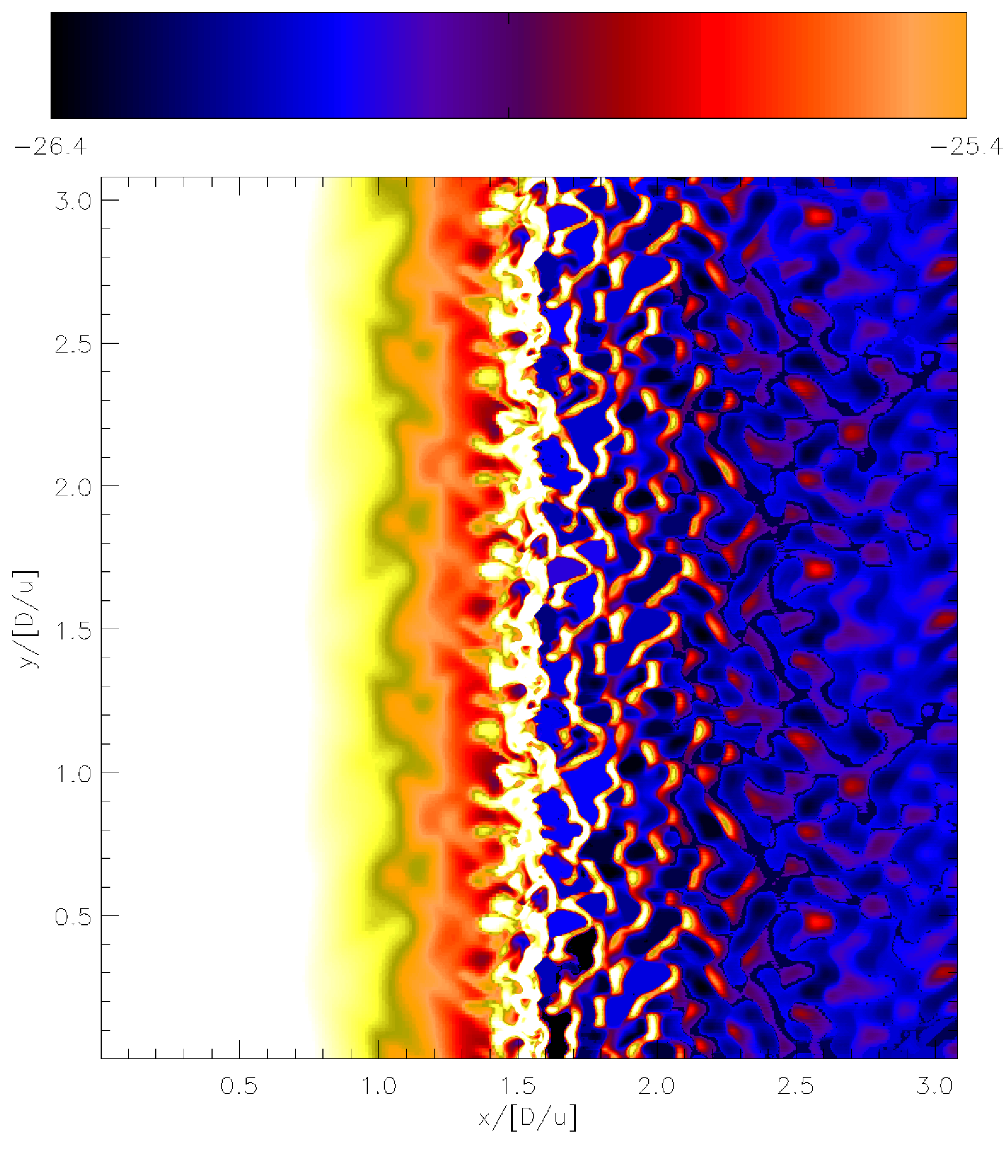}
\includegraphics[width=0.35\textwidth]{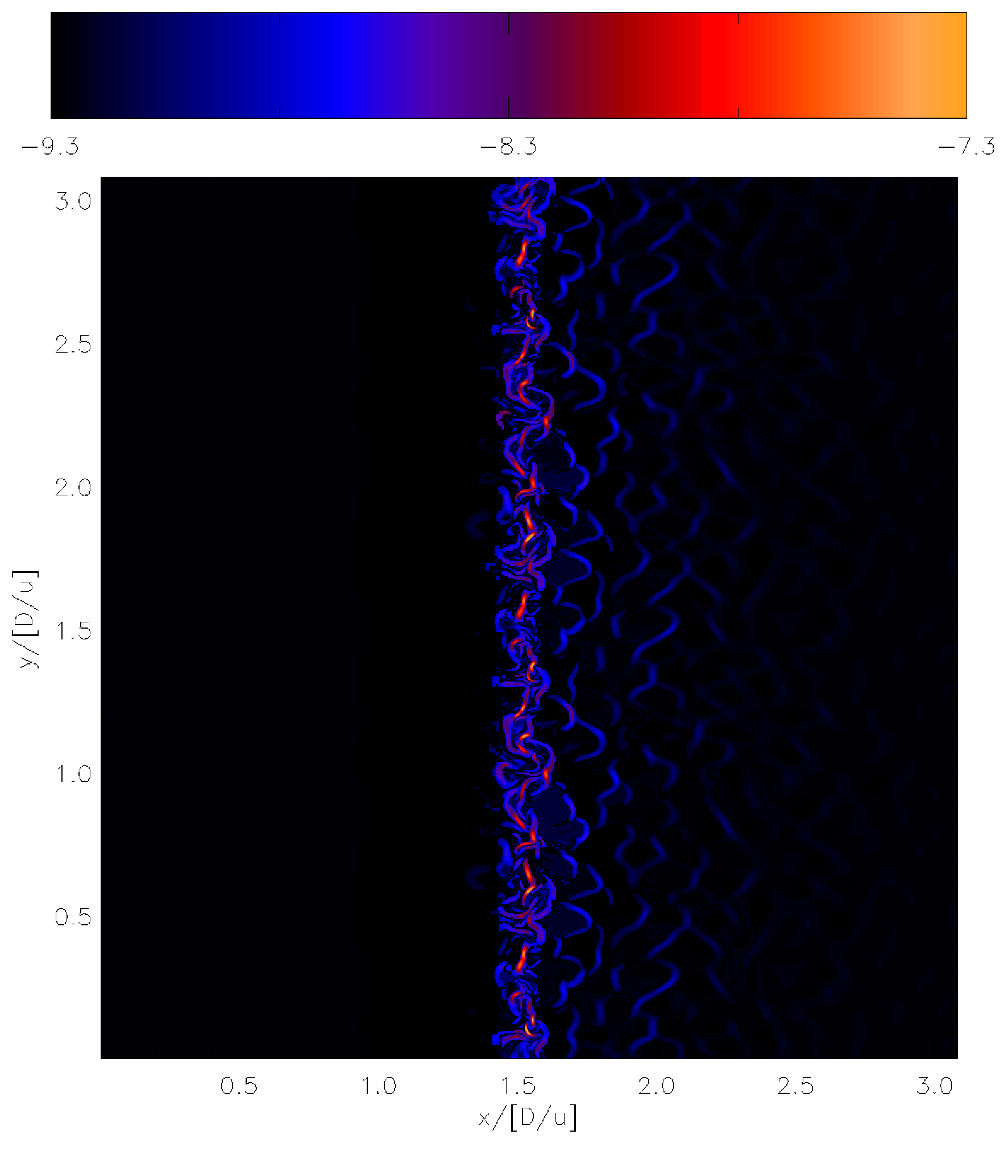}\\
\includegraphics[width=0.35\textwidth]{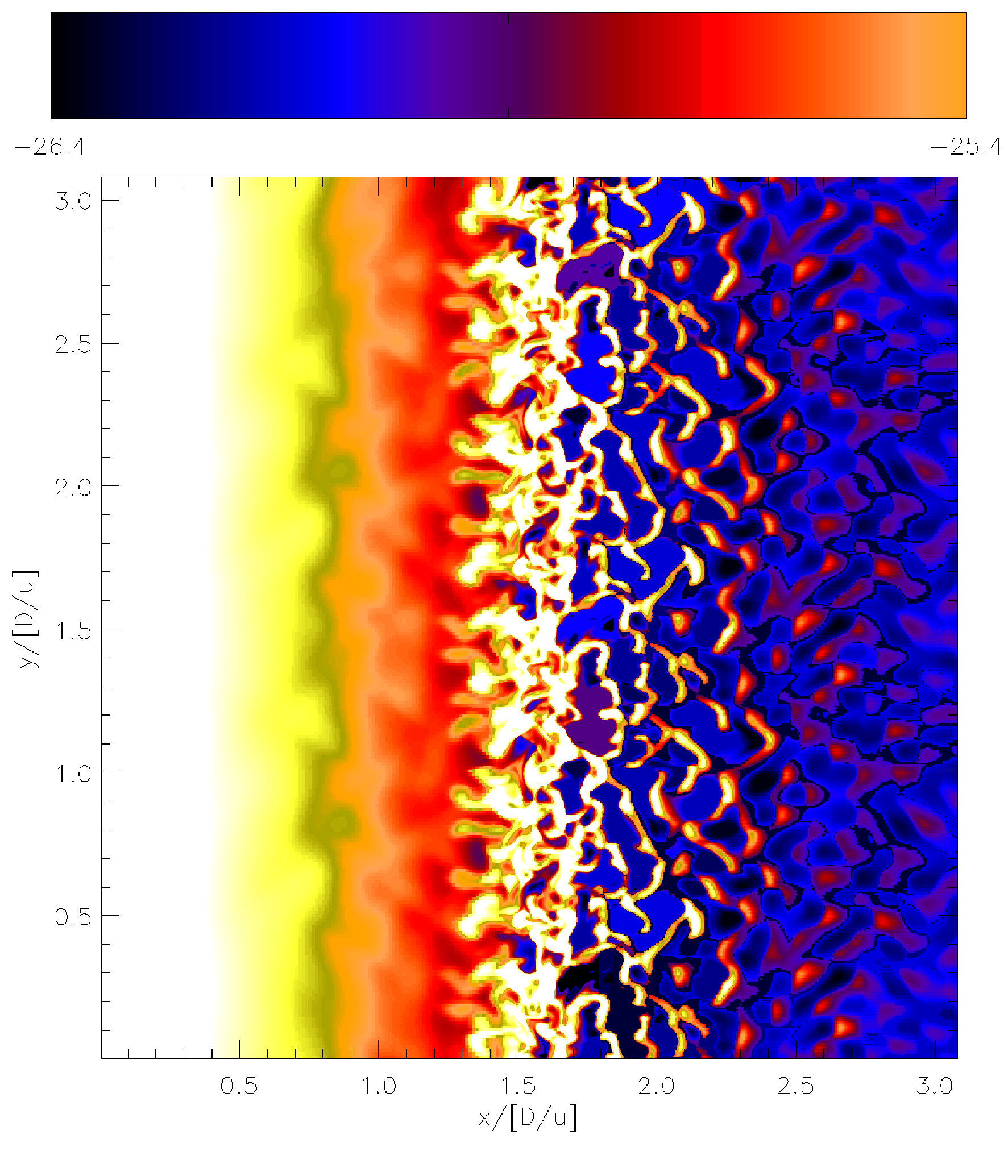}
\includegraphics[width=0.35\textwidth]{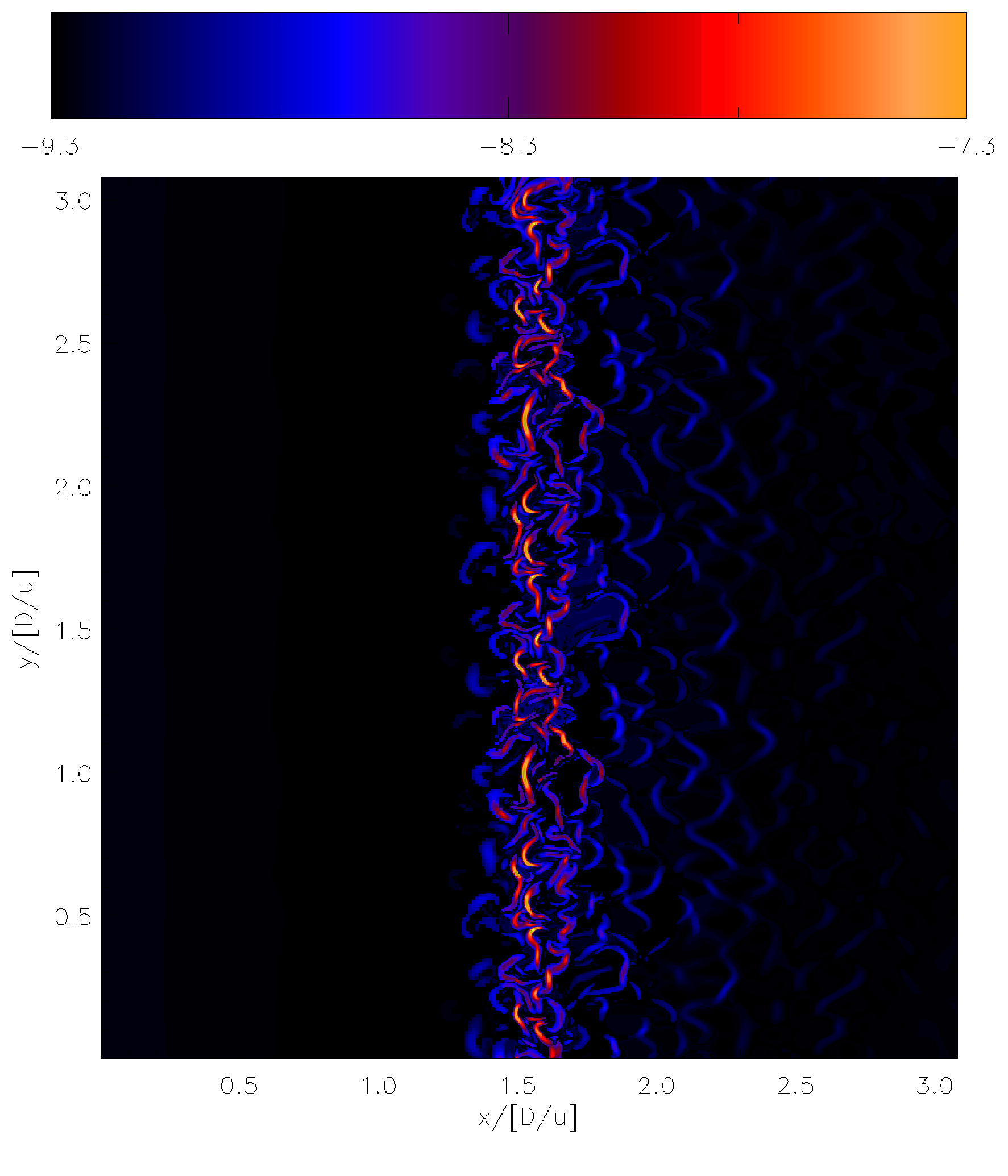}
\caption{Value of gas density (left) and magnetic field strength (right) for three different snapshots (0.16, 0.32 and 0.48 code units of time) for 2D\_M100. The fluid flows from right to left.}
\label{fig:Bdens2D}
\end{center}
\end{figure*}

\begin{figure*}
\begin{center}
\includegraphics[width=0.35\textwidth]{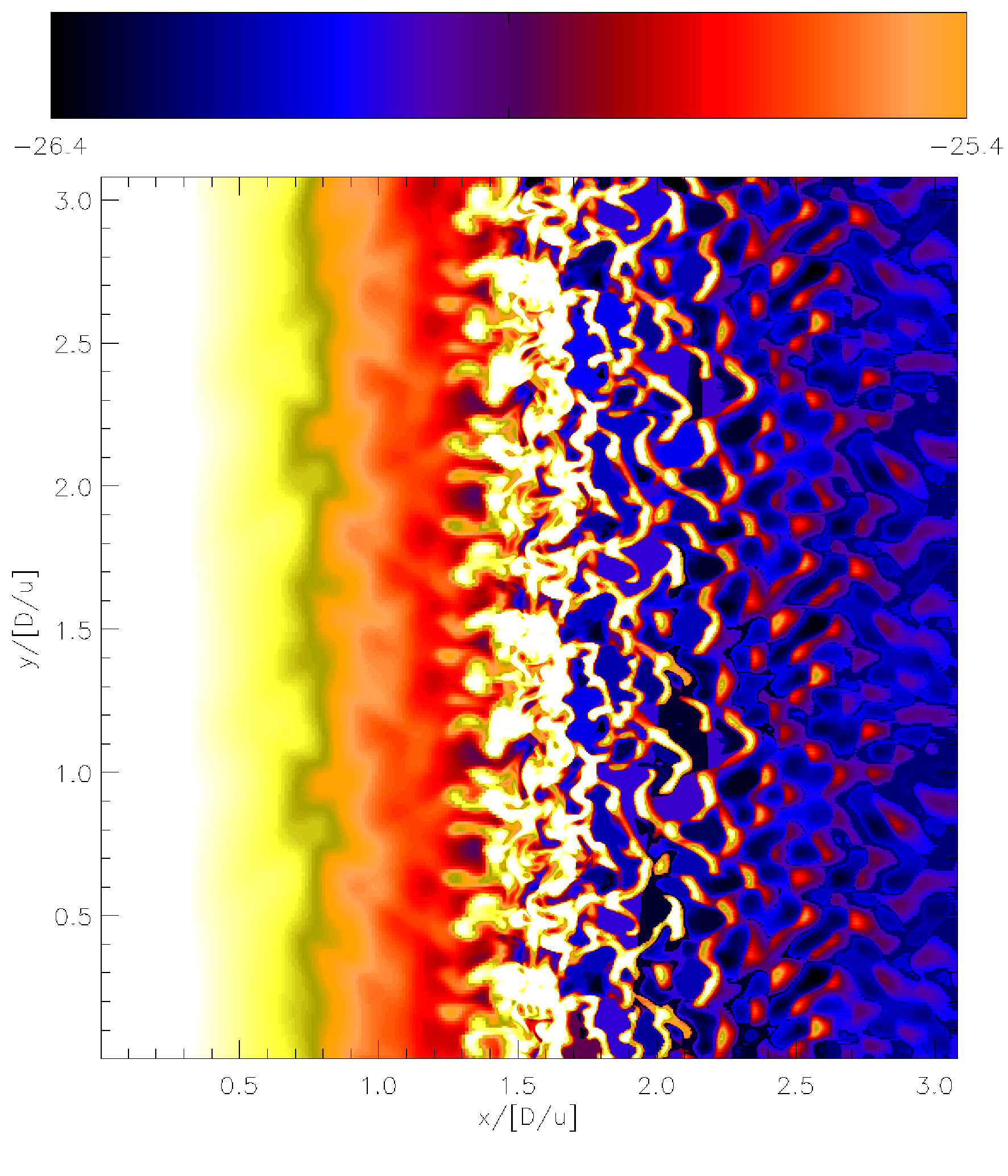}
\includegraphics[width=0.35\textwidth]{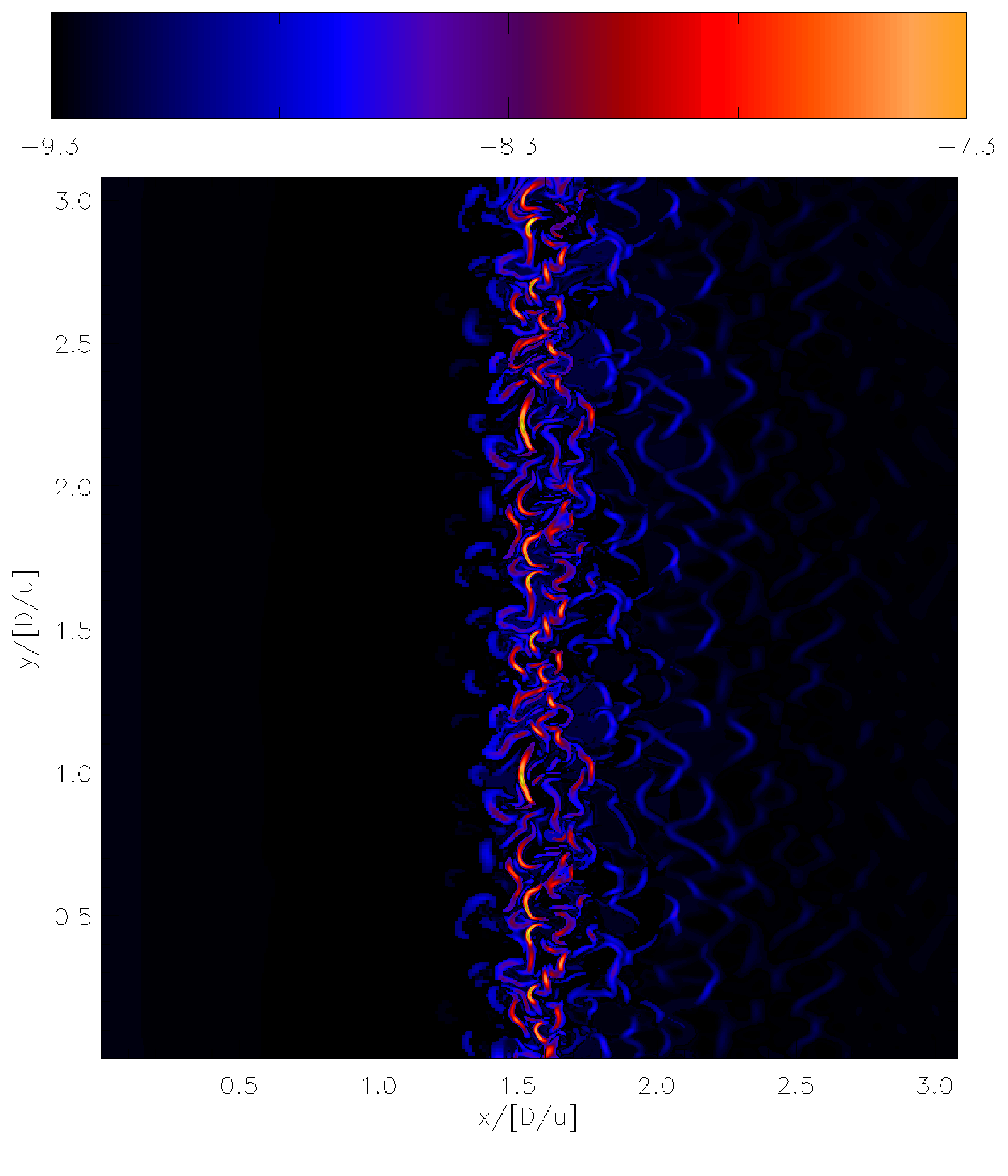}\\
\includegraphics[width=0.35\textwidth]{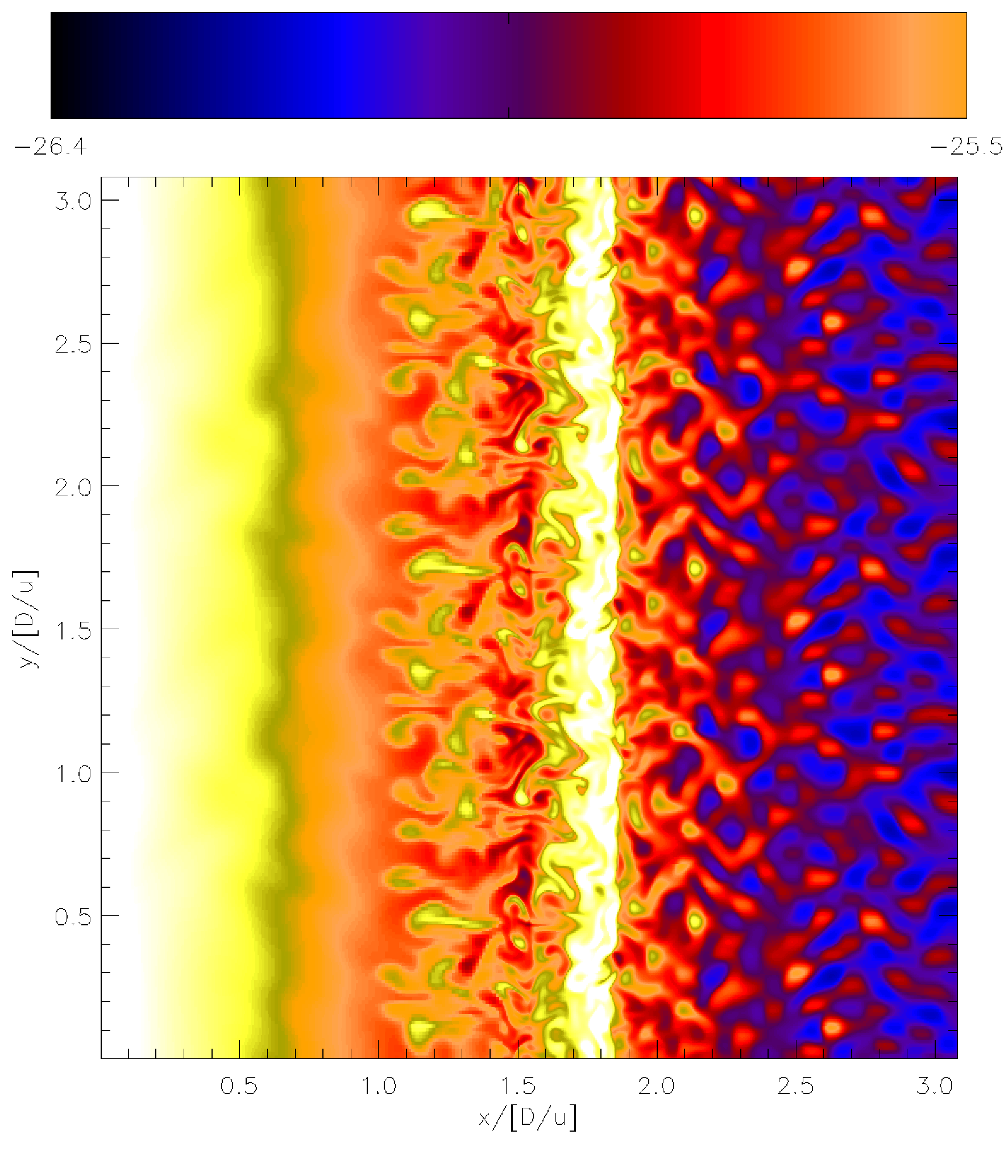}
\includegraphics[width=0.35\textwidth]{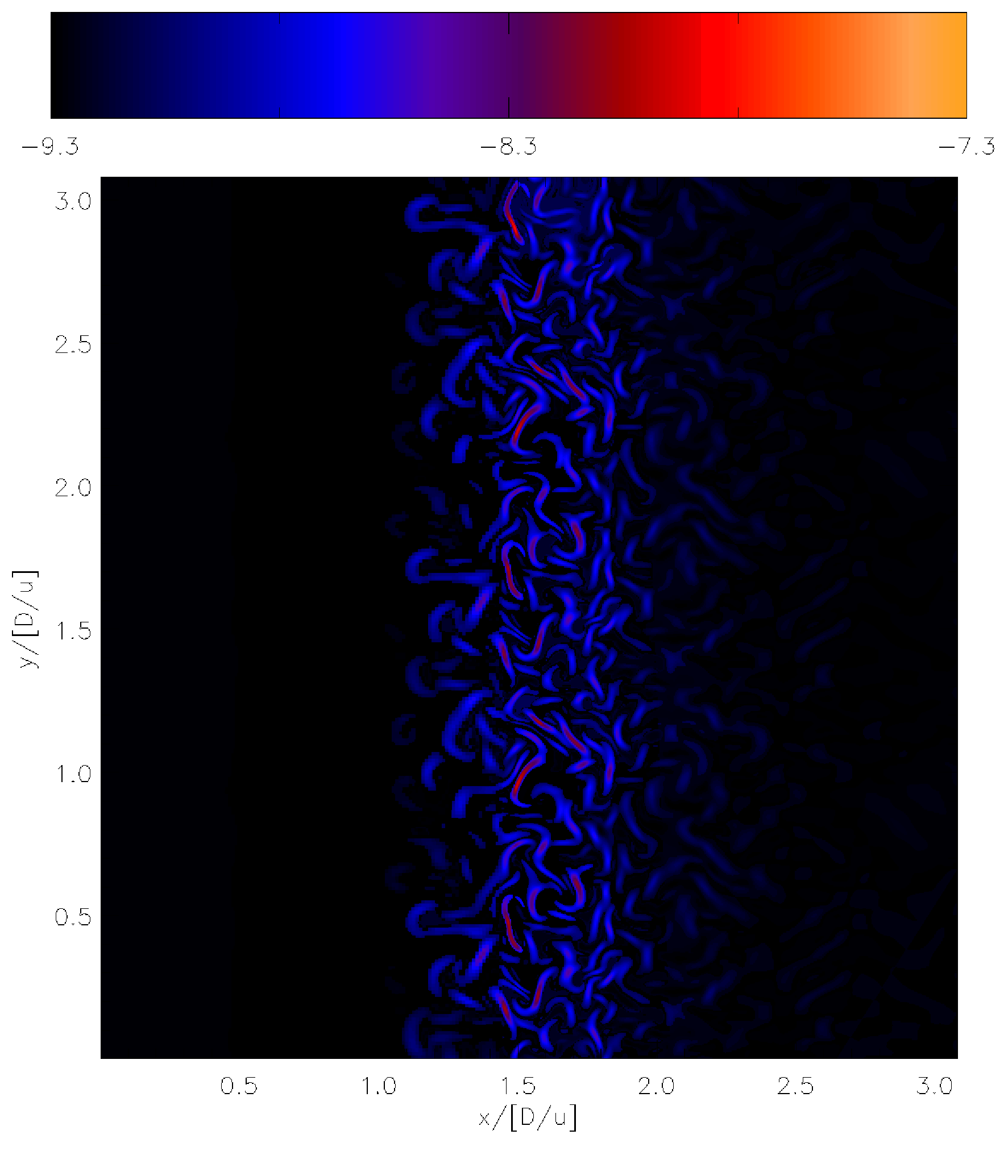}\\
\includegraphics[width=0.35\textwidth]{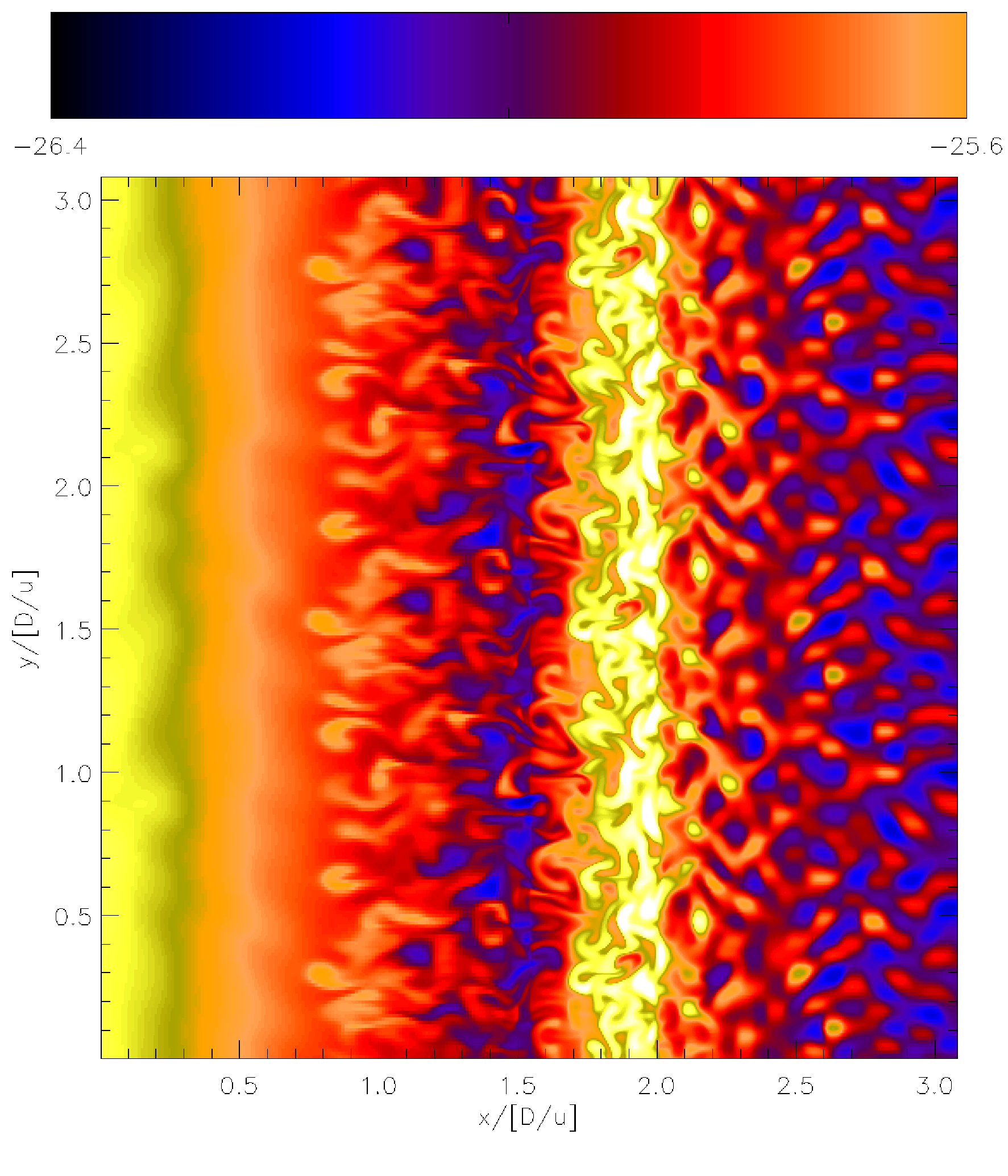}
\includegraphics[width=0.35\textwidth]{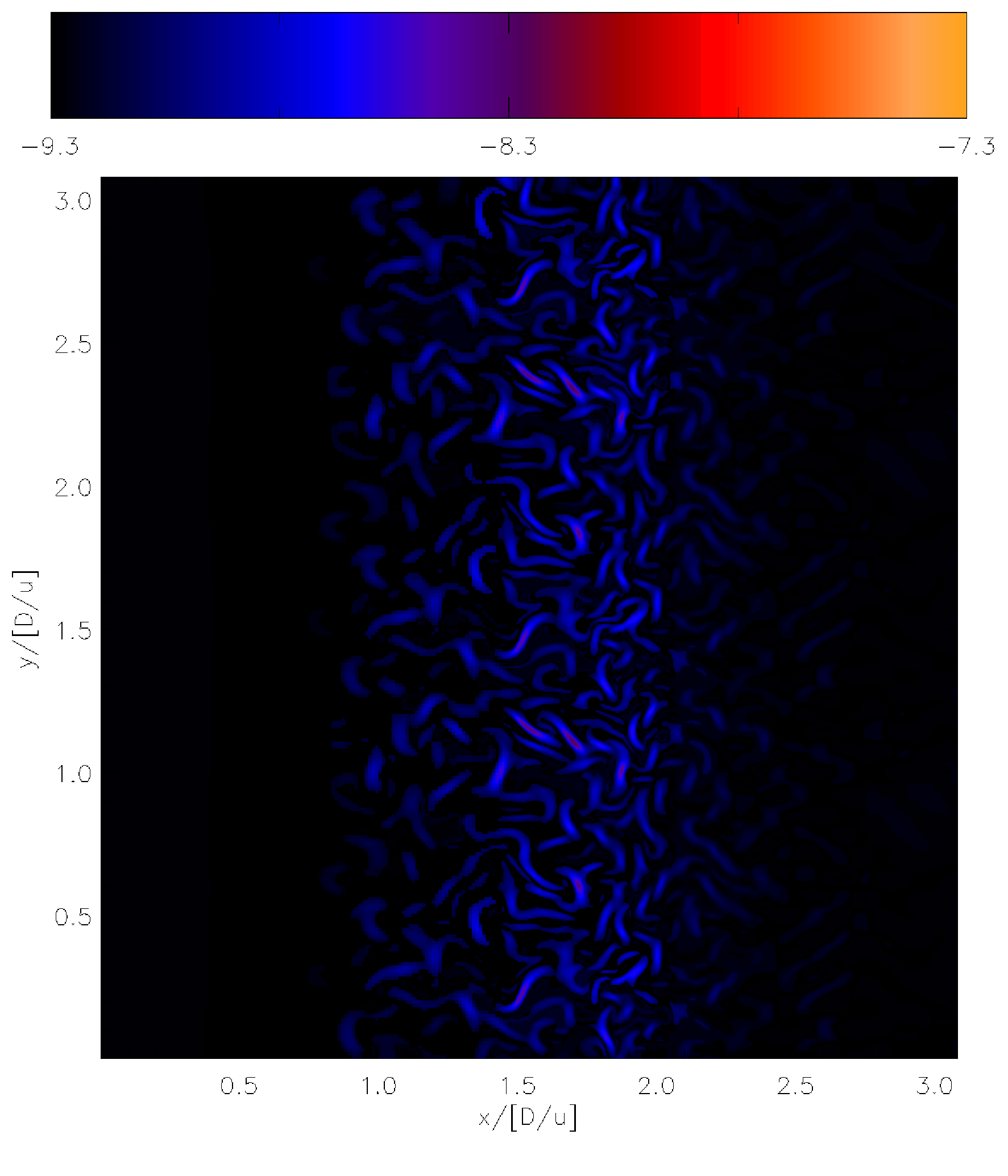}
\caption{Value of gas density (left) and magnetic field strength (right) for (from top to bottom) $M=100$, $M=3$, $M=2$ at $t=0.51$ time units in a 2D simulation.}
\label{fig:dtime}
\end{center}
\end{figure*}

\begin{figure}
\begin{center}
\includegraphics[width=\columnwidth]{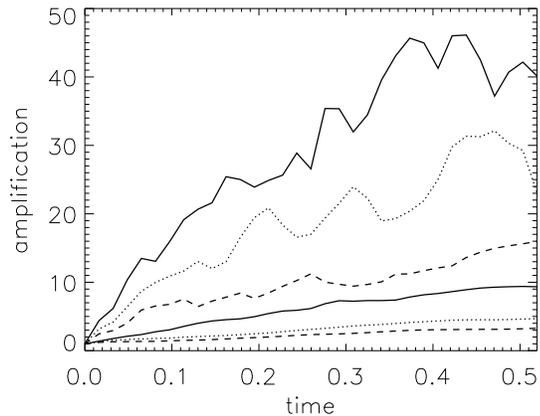}
\caption{Magnetic field amplification as function of time for M=100 (solid), M=3 (dotted), M=2 (dashed). The top curves represent the maximal magnetic field strengths and the lower curves the density-weighted average fields in the upstream region.}
\label{fig:bmax}
\end{center}
\end{figure}

\begin{figure}
\begin{center}
\includegraphics[width=\columnwidth]{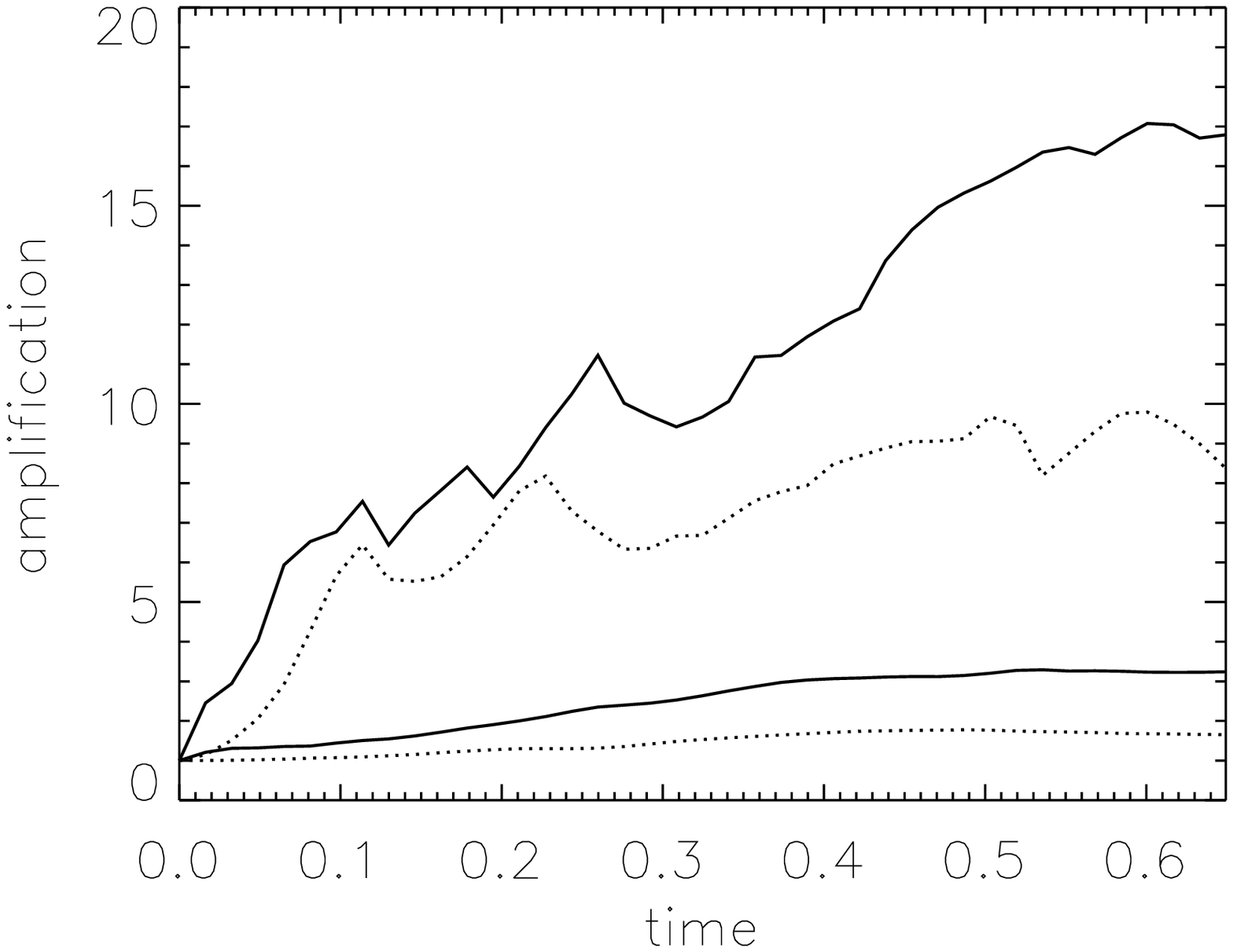}
\caption{Magnetic field amplification as function of time for $M=2$ parallel (dashed) and perpendicular shocks (solid). The top curves represent the maximal magnetic field strengths and the lower curves the density-weighted average fields in the upstream region.}
\label{fig:bmaxperp}
\end{center}
\end{figure}

Fig.~\ref{fig:Bdens2D}  shows snapshots of the gas density and magnetic field strength for three different times for a shock with Mach number $M=100$ shock. The fluid flows from the right to the left. At around 0.3 code units of time after the start of the simulation, the cosmic ray pressure gradient has already amplified the density fluctuations in the upstream region. The magnetic field is strongly amplified by the vorticity generated in the fluid caused by the cosmic-ray pressure. One can clearly observe how the density clumps become flattened parallel to shock surface and break up into smaller pieces as they approach the shock, which is stationary in the middle of the figure panels (note that not the entire computational domain is shown in these images). The magnetic field lines become aligned parallel to shock surface, perpendicular to shock normal. Fig.~\ref{fig:dtime} shows slices of gas density and magnetic field strength for Mach numbers 2, 3 and 100. The amplification as a factor of time for Mach numbers 2, 3 and 100 is shown in Fig.~\ref{fig:bmax}. Both, the maximum amplification in the entire domain (upper lines) as well as the amplification of the density-weighted average field in the immediate upstream region are shown (lower lines).  Here the density-weighted average refers to the $<B>_\rho = \int B\,\rho d^3x/\int \rho d^3x$. The synchrotron emission-weighted magnetic field values will lie somewhere between the maximal and density averaged values, depending on the spectral index (see Sec. 3.3). Clearly, the amplification varies with Mach number. After a significant time, the maximum amplifications achieved are $\sim 40$, $25$ and $15$ for Mach numbers of 100, 3 and 2, respectively. The average magnetic field amplification in the pre-shock region is much lower, though. In a thin strip of width 0.5 $D/u_{\rm up} \eta$, these values are $\sim 9$, $5$ and $3$ for Mach numbers of 100, 3 and 2, respectively. The amplification also depends on the direction of the initial magnetic field. For a parallel shock, i.e. the magnetic field is parallel to the shock normal, the amplification is lower as shown in Fig.~\ref{fig:bmaxperp}.
Clearly, the cosmic ray pressure modifies the shock structure (see e.g. \citet{blandford80, drury81, berezhko99}). In particular, the pressure gradients in the shock precursor enhance the total compression through the shock transition. The total compression through a high Mach number shock can thus greatly exceed the canonical value for strong gasdynamical shocks. In addition, there is the growing magnetic field pressure that also modifies the shock structure and leads to very non-steady complex shock structures. In Fig.~\ref{fig:pprof}, we show profiles (averaged over $y$) of gas pressure, cosmic-ray pressure and magnetic pressure for a $M=100$ shock around the shock transition. A study on the shock modification in the presence of this mode of magnetic field amplification will be the subject of a separate study.

\subsection{Magnetic field power spectra and Faraday rotation maps}

\begin{figure}
\begin{center}
\includegraphics[width=\columnwidth]{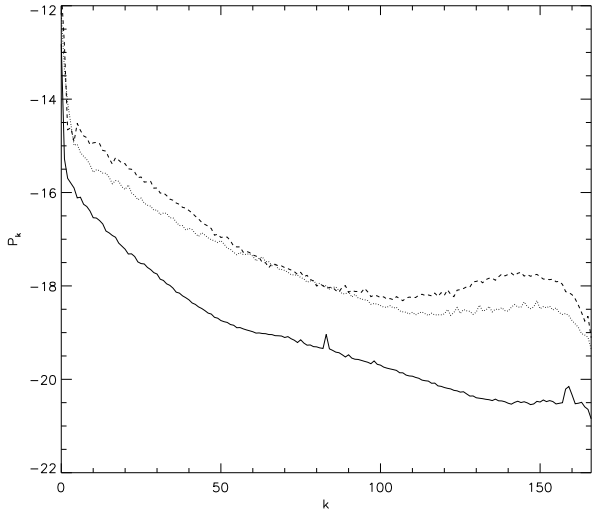}
\caption{Power spectra for the total magnetic field for the 2D\_M100 run for 0.07 (solid), 0.32 (dotted), 0.64 (dashed) time units. The wavenumber $k$ is in code units.}
\label{fig:bpower}
\end{center}
\end{figure}

In Fig.~\ref{fig:bpower} we show the power spectrum of the magnetic field that is computed in a rectangle centred on the shock front.  The power spectrum shows that small modes are preferentially amplified as the power at larger wave numbers $k$ increases. This is because on an advection time-scale smaller pockets of lower density suffer greater deceleration and thus smaller perturbation scales get amplified more, as also seen from equation (\ref{eqn:lambda-limit}).\\ 

\begin{figure}
\begin{center}
\includegraphics[width=\columnwidth]{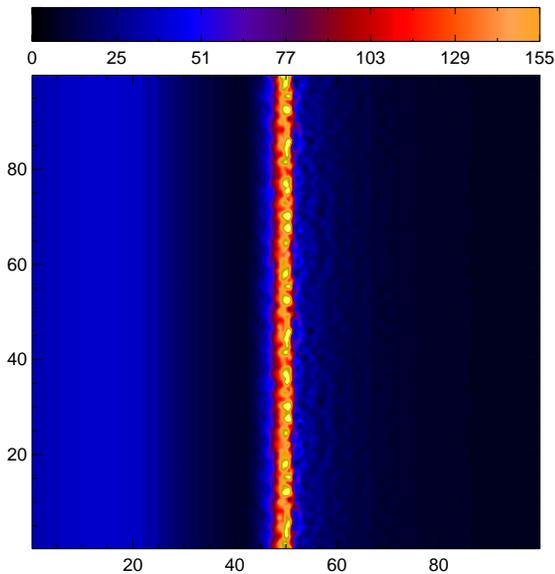}
\caption{Faraday map for a 3D run with $M=100$ at $t=0.48$ time units, projected along the $y$-direction. Here units typical for galaxy clusters were chosen by equating the unit of length to 100 kpc, unit of velocity to $10^8$ cm s$^{-1}$ and unit of density to $10^{-26}$ g cm$^{-3}$.}
\label{fig:faraday}
\end{center}
\end{figure}

The most promising way in which the topology of the magnetic field can be observed is via maps of the Faraday rotation.
Synchrotron radiation (from background sources) propagating through a magnetised plasma undergoes Faraday rotation, and the rotation measure RM is given by:
\begin{equation}
\label{rm}
{\rm RM} = 812 \int _0 ^{L_{\rm path}} n_{\rm e} B_{\parallel} dl\ {\rm rad\ m^{-2}}
\end{equation}
where $L_{\rm path}$ is the path along the line of sight in units of ${\rm kpc}$, $n_{\rm e} = \rho / (m_{\rm p} \mu_{\rm e})$ is the electron density in ${\rm cm^{-3}}$ ($m_{\rm p}$ is the proton mass, $\mu_{\rm e} \simeq 1.14$ is the average mass per electron in a.m.u), and $B_{\parallel}$ is the component of $B$ along the line of sight, in $\mu{\rm G}$. Approximating the quantities in the integrand as constants along the line of sight, one gets:
\begin{equation}
\label{rm1}
{\rm RM} = 812 \left (\frac{n_{\rm e}}{{\rm cm^{-3}}} \right ) \left (\frac{B_{\parallel}}{\mu{\rm G}} \right ) \left (\frac{L_{\rm path}}{{\rm kpc}} \right ) {\rm rad\ m^{-2}}
\end{equation}
For the 3D run with a Mach number of $M=100$ a Faraday map, projected along the $y$-direction, is shown in Fig.~\ref{fig:faraday}. The RM fluctuates on small scales and extends into the precursor. The RM precursor can be probed with sensitive wide-band radio telescopes. Here the most promising technique is rotation measure synthesis. The application of a Fourier transform to spectropolarimetric data in radio continuum, Faraday rotation measure synthesis, yields the ÒFaraday spectrumÓ, which contains information about the magneto-ionic medium along the line of sight. The range of scales probed is related to the width in wavelength space that is covered by the observation. The planned Square Kilometre Array (SKA) will be the ideal instrument for this \citep{gaensler08}.  Moreover, one can probe the magnetic field with background sources. The SKA Magnetism Key Science Project plans to survey parts of the sky at around 1 GHz with 1 h integration per field which should detect sources of 0.5-1 $\mu$Jy flux density and measure at least 1500 RMs deg$^{-2}$, or a total of at least $1.5\times10^7$ RMs from compact polarized extragalactic sources at a mean spacing of $\sim 90''$. For a relic such as the sausage relic in CIZA J2242.8+5301 that spans 9' on the sky, one can probe variations on scales of 1.7 Mpc/ (9'/90'') $\sim$ 280 kpc.

\subsection{Synchrotron precursors}

Cosmic-ray electrons that diffuse ahead of the shock can produce synchrotron radiation in the upstream magnetic field. The synchrotron emission from the shock precursor or the lack thereof can thus be used to validate the CR-induced magnetic field amplification. The cosmic ray pressure is dominated by the cosmic-ray protons and is computed in our simulations as a function of position. Using DSA theory, we can deduce the density of CR electrons. 


The differential number density distribution of a CR population can be
described by a power-law in momentum $p$,

\begin{equation} 
f(p)=\frac{{\rm d}N}{{\rm d}p\,{\rm d}V} = C p^{-\alpha}\,\theta(p-q) , 
\end{equation} 
where $C$ gives the normalization, $q$ is a low-momentum cut-off, and $\alpha$ is the power-law index $\alpha={r+2 \over r -1}$ assumed to be the same for protons and electrons, and $r$ is the compression ratio. The momenta are expressed in dimensionless form in units of $m_{\rm p}c$, where $m_{\rm p}$ is the proton mass. In order to determine the compression ratio for the computation of the cosmic-ray pressure, we have used the values of the density 5 cells up-, and 5 cell downstream of the shock position that was found at run-time.

The pressure of this cosmic-ray population is given by 
\begin{equation} 
P_{\rm CR} = \frac{C\, m_{\rm p} c^2\,}{6} \, {\cal B}_{\frac{1}{1+q^2}} \left( \frac{\alpha-2}{2},\frac{3-\alpha}{2} \right) , 
\label{pcr}
\end{equation}
while the number density is simply $n_{\rm CR} = {C\, q^{1-\alpha}}/({\alpha-1})$.  Here 

\begin{equation} 
{\cal B}_n(a,b)\equiv \int_0^n x^{a-1} (1-x)^{b-1}\,{\rm  d}x 
\end{equation} 
denotes the incomplete Beta function. The beta function does not vary much for $\alpha < 2.5$ and $q\leq1$. Here we choose $q=1$.

Using equation (\ref{pcr}) and the expression for $n_{\rm CR}$, the ratio of CR-pressure to shock ram pressure can be written as

\begin{equation}
\frac{p_{\rm CR}}{p_{\rm ram}}=\frac{n_{\rm CRp} m_pc^2q^{\alpha-1}(\alpha-1){\cal B}_{\frac{1}{1+q^2}} \left( \frac{\alpha-2}{2},\frac{3-\alpha}{2} \right)}{6\rho_{\rm up} u_{\rm up}^2} \ .
\end{equation}

Ignoring losses, the normalization of the electron distribution function is obtained from \citep{ellison00}:
\begin{equation}
n_{\rm CRe}  =n_{\rm CRp}  {\eta_{\rm inj}^e \over \eta_{\rm inj}^p} \bigg({m_e \over m_p}\bigg)^{(\alpha-3)/2} ~,
\label{eq10}
\end{equation}
where $\eta_{\rm inj}^e$($\eta_{\rm inj}^p$) is the electron (proton) injection parameter (i.e. the fraction of shocked electrons with superthermal energies). The parameter $\eta_{\rm inj}$ is poorly constrained and here we use  $\eta_{\rm inj}^e=\eta_{\rm inj}^p=10^{-4}$.

The radio emissivity $j_\nu$ at frequency $\nu$ and per steradian is given by \citep{rybicki86}
\begin{equation}
   j_\nu(\mathbf{x})=c_2(\alpha)\, n_{\rm CRe} \, 
   B_\perp^{\frac{\alpha+1}{2}}
   \left(\frac{\nu}{c_1}\right)^{-\frac{\alpha -1}{2}}
\end{equation}
with $c_1=3\,e\,{\rm GeV^2}/(2\,\pi\, m_\e^3\,c^5)\,$,
\begin{equation}
   c_2(\alpha)=\frac{\sqrt{3}}{16\pi}\frac{e^3}{m_\mathbf{e}c^2}
               \frac{\alpha+\frac{7}{3}}{\alpha+1}
               \Gamma\left( \frac{3\alpha-1}{12}\right)
               \Gamma\left( \frac{3\alpha+7}{12}\right),
\end{equation}
where $\Gamma$ the Gamma function and $B_\perp$  the magnetic field component within the plane of the sky. 

\begin{figure}
\begin{center}
\includegraphics[width=\columnwidth]{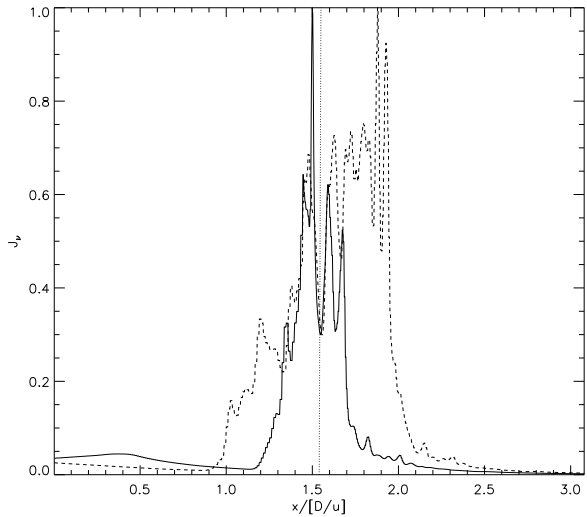}
\caption{Synchrotron emissivity profiles (normalised to maximal emissivity) across the shock for $M=3$ at 0.32 time units (solid) and 0.64 time units (dashed). The vertical dashed line marks the location of the shock.}
\label{fig:synch}
\end{center}
\end{figure}

Using the magnetic field as well as the cosmic ray pressure from our simulation output, we can compute the synchrotron emissivity. This is shown in Fig.~\ref{fig:synch} for times 0.32 and 0.64 time units after the start of the simulation. Clearly, as the upstream magnetic field increases, so does the synchrotron precursor, and the synchrotron precursor's extent into the upstream region increases with time. After about 0.64 time units the synchrotron precursor extends 0.5 $D/u_{\rm up}$ ahead of the shock. For the case of galaxy cluster outskirts this corresponds to a width

\begin{equation}
w\sim16\, {\rm kpc}\, \left ( \frac{D}{10^{31} {\rm cm}^2 {\rm s}^{-1}} \right ) \left (\frac{u_{\rm up}}{10^8 {\rm cm\, s}^{-1}} \right )^{-1} .  
\end{equation}
Since the position of the shock as determined by X-ray observations is at best only known to within 50 kpc, only a limited range of parameters can be probed. In SNRs for $D=10^{26} {\rm cm}^2 {\rm s}^{-1}$ and velocities of $u=10^9 {\rm cm\, s}^{-1}$, $w=5\times 10^{14}$ m. Here we have neglected radiative losses, such as Inverse Compton and synchrotron losses. Including these losses will lead to a steepening of the radio spectrum away from the shock, as discussed, e.g. in \citep{hoeft07}. The Low Frequency Array (LOFAR) operates at low frequencies (30-240 MHz) and still has arcsecond resolution once the international baselines are included. Synthetic radio observations of shock precursors that include radiative losses and instrumental effects are the subject of future work.

\subsection{Dependence on resolution}

\begin{figure}
\begin{center}
\includegraphics[width=\columnwidth]{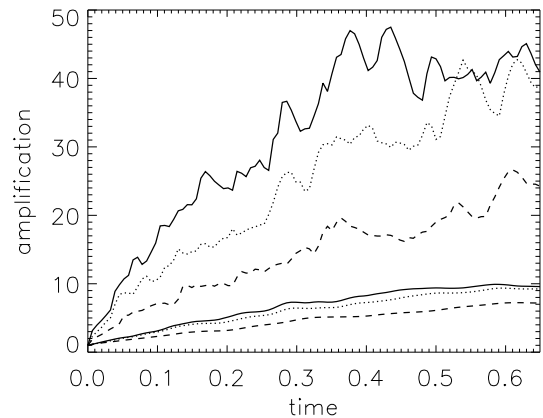}
\caption{Magnetic field amplification as function of time for $M=100$ for effective resolutions $2048\times 1024$ (solid), $1024\times 512$ (dotted) and $512\times 256$ (dashed). The top curves represent the amplification of maximal magnetic field strengths and the lower curves the density-weighted average field strengths in the upstream region.}
\label{fig:bmaxres}
\end{center}
\end{figure}

The amplification of the magnetic field as a function of time for different resolutions is plotted in Fig.~\ref{fig:bmaxres}. It appears that the magnetic field amplification is reasonably converged, even at the relatively high resolution. Higher resolution results in greater magnetic field amplification as the field can be amplified on a greater range of length scales, and hence may be subject to higher amplification. We should note that the length scales on which the turbulence is driven are those on which the density varies. This is very different to the modelling of driven turbulence turbulence where the driving scales are usually large, allowing the energy to cascade down to the dissipation length scale. While in this case at least part of the inertial range can be simulated, in our case there is inherently no inertial range. So in principle, one would expect that the magnetic field values found in our simulations represent lower limits to the real amplification that may occur.

Indeed, we find that the amplification is significantly higher between the $1024\times 512$ and $512\times 256$ runs. However, Fig.~\ref{fig:bmaxres} shows that there is hardly any difference, both, in maximum as well as density-weighted averaged field between the $2048\times 1024$ and $1024\times 512$ runs.

\begin{figure}
\begin{center}
\includegraphics[width=\columnwidth]{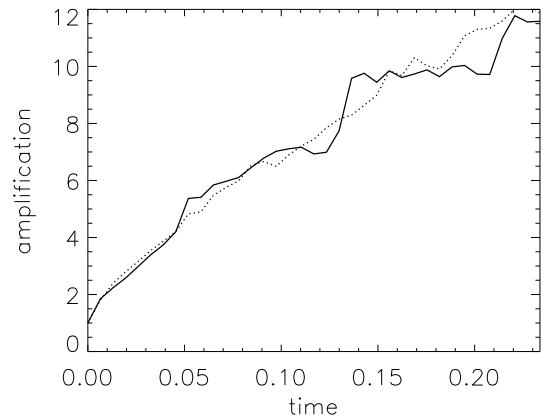}
\caption{Comparison of amplification factor as a function of time in 2D (solid) and 3D (dashed) of the same spatial resolution.}
\label{fig:d2d3d}
\end{center}
\end{figure}

\begin{figure}
\begin{center}
\includegraphics[width=0.35\textwidth]{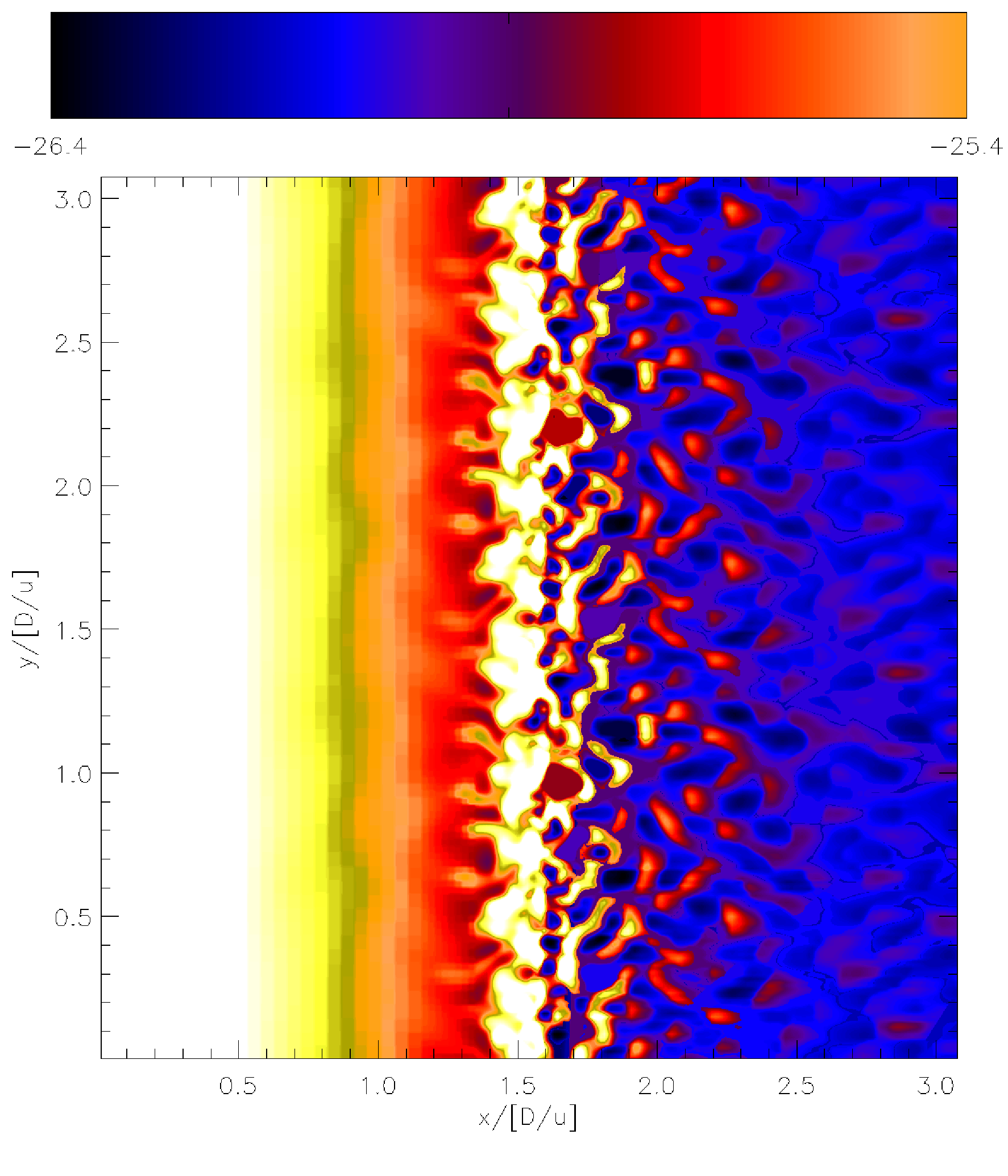}
\includegraphics[width=0.35\textwidth]{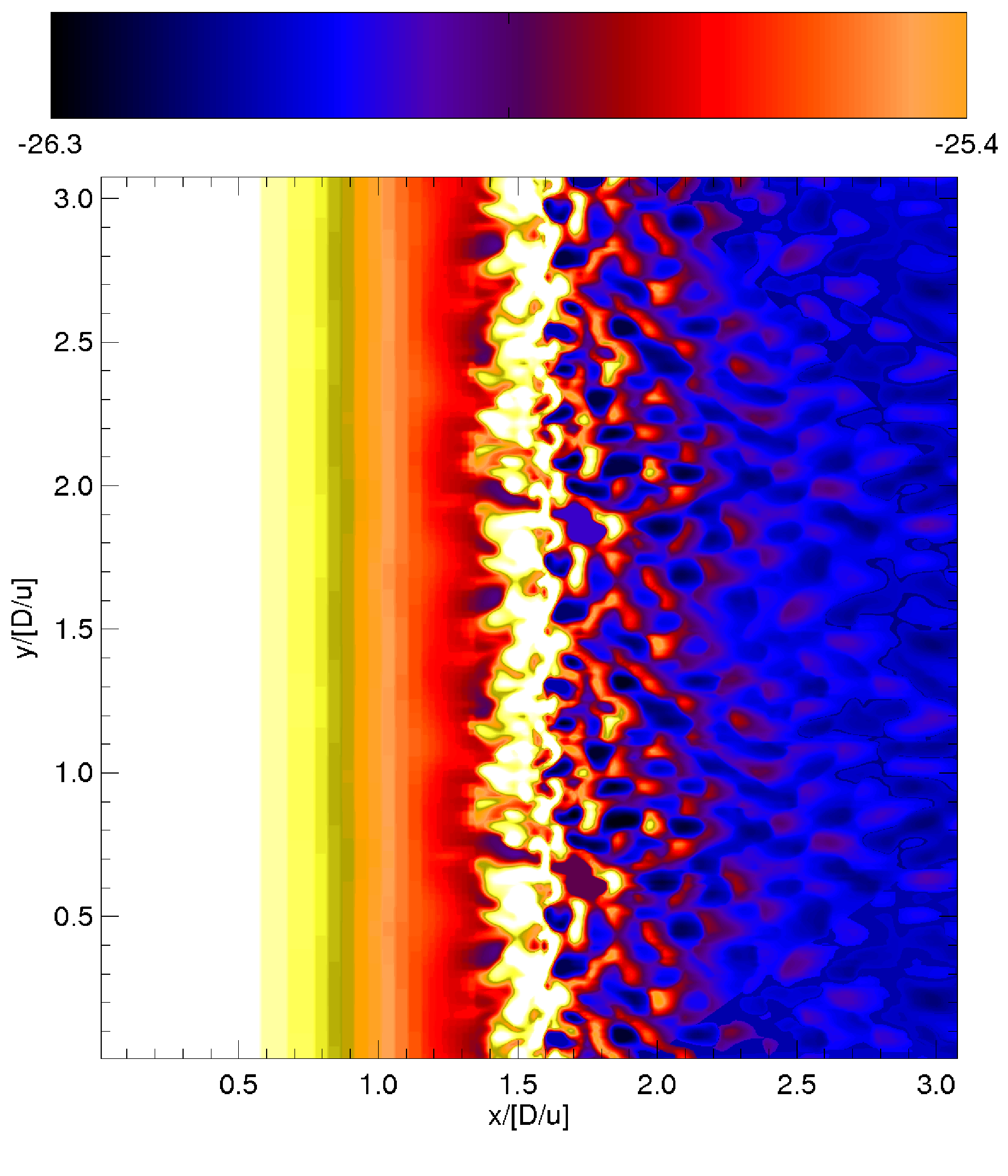}
\caption{Comparison of gas density in 2D (top) and 3D (bottom) runs at time 0.42 code units.}
\label{fig:d2d3d_2}
\end{center}
\end{figure}

It is frequently found that 2D and 3D turbulence are different \citep{biskamp03}. However, in this particular case, we find little difference. In Fig.~\ref{fig:d2d3d}, we show that the maximum magnetic field in 2D and 3D simulations evolve very similarly. The structure of the density and the total magnetic field is also similar as seen in Fig.~\ref{fig:d2d3d_2}. Given that the 3D simulations are computationally very expensive, this means that significant progress can be made with 2D simulations.

\section{Conclusions}

Without the effects of cosmic-ray pressure, magnetic fields experience little amplification during shock passage. In their MHD simulations of shock-cloud interaction, \cite{shin08} find that after the passage of the shock through a cloudy medium the magnetic field is amplified by a factor that depends on the orientation of $B$ and on the $\beta$ parameter, defined as the ratio between the thermodynamical and magnetic pressure. As worked out in \cite{iapichino12}, without CRs, a shock passing through a clumpy intracluster medium would lead to an amplification of no more than a few percent. In this paper, we included the effects of cosmic-ray pressure on upstream density inhomogeneities. We performed 2D and 3D simulations of shocks that include the diffusion of CRs away from their injection point at the shock and the interaction of the CRs with an inhomogeneous upstream medium. In summary, we find:

\begin{itemize}
\item For high Mach number shocks, such as those found in SNRs, the amplification by cosmic-ray driven turbulence is significant. For a $M=100$ shock, the maximum field can increase by a factor of $\sim 40$. Thus, our 3D simulations confirm the efficiency of his mechanism that has been found in previous work and it is likely that this process contributes to the magnetic field amplification in supernova remnants. In weak ($M\sim 2-3$) shocks, cosmic-ray driven turbulence can amplify the maximum magnetic fields by a factor of 10-20 and by a factor 3-5 in the density-weighted average. In order to achieve fields of around 5 $\mu$G in radio relics, the field in the inflowing fluid still needs to have a strength of $\sim 1\, \mu$G assuming an efficient conversion of kinetic energy to CR energy ($\eta=0.1$).  

\item In cluster merger shocks one would expect synchrotron precursors of width 0.5 $D/u_{\rm up}$ which for $D=10^{31}$ cm$^2$s$^{-1}$ and $u_{\rm up}=10^8$ cm s$^{-1}$ is 

\begin{equation}
w\sim 16\, {\rm kpc}\, \left ( \frac{D}{10^{31} {\rm cm}^2 {\rm s}^{-1}} \right ) \left (\frac{u_{\rm up}}{10^8 {\rm cm\, s}^{-1}} \right )^{-1} .  
\end{equation}

Provided one can locate the shock to that precision using X-ray observations, one can set limits on $D$ and the properties of upstream density perturbations.

\item The cosmic-ray driven instability preferentially amplifies small perturbations. This leads to Faraday rotation maps that vary on the scale set by the initial density perturbations. 
The RM fluctuates on small scales and extends into the precursor. With sensitive wide-band radio telescopes, the RM precursor can be probed and this mechanism of cosmic-ray driven turbulence be verified.

\end{itemize}

\section*{Acknowledgements}

MB acknowledges support by the research group FOR 1254 funded by the Deutsche Forschungsgemeinschaft. The results presented were produced using the FLASH code, a product of the DOE ASC/Alliances-funded Center for Astrophysical Thermonuclear Flashes at the University of Chicago. Simulations were performed on the {\sc JUROPA} supercomputer at the Forschungszentrum J\"ulich under grant NIC 5984 an 5056. Dr K.M. Schure is acknowledged for inspiring conversations. Dr Annalisa Bonafede and Dr Franco Vazza are thanked for valuable input, and an anonymous referee is thanked for spotting several mistakes in the original manuscript.

\bibliographystyle{mn2e}
\bibliography{Bamplification}

\begin{thebibliography}{}

\bibitem[\protect\citeauthoryear{{Axford}, {Leer} \& {Skadron}}{{Axford}
  et~al.}{1977}]{1977ICRC...11..132A}
{Axford} W.~I.,  {Leer} E.,    {Skadron} G.,  1977, in International Cosmic Ray
  Conference Vol.~11 of International Cosmic Ray Conference, {The Acceleration
  of Cosmic Rays by Shock Waves}.
pp 132--+

\bibitem[\protect\citeauthoryear{{Bell}}{{Bell}}{1978}]{1978MNRAS.182..147B}
{Bell} A.~R.,  1978, \mnras, 182, 147

\bibitem[\protect\citeauthoryear{{Bell}}{{Bell}}{2004}]{bell04}
{Bell} A.~R.,  2004, \mnras, 353, 550

\bibitem[\protect\citeauthoryear{{Bell}, {Schure}, {Reville} \&
  {Giacinti}}{{Bell} et~al.}{2013}]{bell13}
{Bell} A.~R.,  {Schure} K.~M.,  {Reville} B.,    {Giacinti} G.,  2013, \mnras,
  431, 415

\bibitem[\protect\citeauthoryear{{Beresnyak}, {Jones} \&
  {Lazarian}}{{Beresnyak} et~al.}{2009}]{beresnyak09}
{Beresnyak} A.,  {Jones} T.~W.,    {Lazarian} A.,  2009, \apj, 707, 1541

\bibitem[\protect\citeauthoryear{{Berezhko} \& {Ellison}}{{Berezhko} \&
  {Ellison}}{1999}]{berezhko99}
{Berezhko} E.~G.,  {Ellison} D.~C.,  1999, \apj, 526, 385

\bibitem[\protect\citeauthoryear{{Biskamp}}{{Biskamp}}{2003}]{biskamp03}
{Biskamp} D.,  2003, {Magnetohydrodynamic Turbulence}

\bibitem[\protect\citeauthoryear{{Blandford}}{{Blandford}}{1980}]{blandford80}
{Blandford} R.~D.,  1980, \apj, 238, 410

\bibitem[\protect\citeauthoryear{{Blandford} \& {Ostriker}}{{Blandford} \&
  {Ostriker}}{1978}]{1978ApJ...221L..29B}
{Blandford} R.~D.,  {Ostriker} J.~P.,  1978, \apjl, 221, L29

\bibitem[\protect\citeauthoryear{{Bonafede}, {Dolag}, {Stasyszyn}, {Murante} \&
  {Borgani}}{{Bonafede} et~al.}{2011}]{bonafede11}
{Bonafede} A.,  {Dolag} K.,  {Stasyszyn} F.,  {Murante} G.,    {Borgani} S.,
  2011, \mnras, 418, 2234

\bibitem[\protect\citeauthoryear{{Br{\"u}ggen}, {Ruszkowski}, {Simionescu},
  {Hoeft} \& {Dalla Vecchia}}{{Br{\"u}ggen} et~al.}{2005}]{bruggen05}
{Br{\"u}ggen} M.,  {Ruszkowski} M.,  {Simionescu} A.,  {Hoeft} M.,    {Dalla
  Vecchia} C.,  2005, \apjl, 631, L21

\bibitem[\protect\citeauthoryear{{Bykov}, {Osipov} \& {Ellison}}{{Bykov}
  et~al.}{2011}]{bykov11}
{Bykov} A.~M.,  {Osipov} S.~M.,    {Ellison} D.~C.,  2011, \mnras, 410, 39

\bibitem[\protect\citeauthoryear{{Caprioli}, {Kang}, {Vladimirov} \&
  {Jones}}{{Caprioli} et~al.}{2010}]{caprioli10}
{Caprioli} D.,  {Kang} H.,  {Vladimirov} A.~E.,    {Jones} T.~W.,  2010,
  \mnras, 407, 1773

\bibitem[\protect\citeauthoryear{{Casse}, {Lemoine} \& {Pelletier}}{{Casse}
  et~al.}{2002}]{casse02}
{Casse} F.,  {Lemoine} M.,    {Pelletier} G.,  2002, \prd, 65, 023002

\bibitem[\protect\citeauthoryear{{Dolag}, {Bartelmann} \& {Lesch}}{{Dolag}
  et~al.}{2002}]{dolag02}
{Dolag} K.,  {Bartelmann} M.,    {Lesch} H.,  2002, \aap, 387, 383

\bibitem[\protect\citeauthoryear{{Drury} \& {Downes}}{{Drury} \&
  {Downes}}{2012}]{drury12}
{Drury} L.~O.,  {Downes} T.~P.,  2012, \mnras, 427, 2308

\bibitem[\protect\citeauthoryear{{Drury} \& {Voelk}}{{Drury} \&
  {Voelk}}{1981}]{drury81}
{Drury} L.~O.,  {Voelk} J.~H.,  1981, \apj, 248, 344

\bibitem[\protect\citeauthoryear{{Eckert}, {Vazza}, {Ettori}, {Molendi},
  {Nagai}, {Lau}, {Roncarelli}, {Rossetti}, {Snowden} \&
  {Gastaldello}}{{Eckert} et~al.}{2012}]{eckert12}
{Eckert} D.,  {Vazza} F.,  {Ettori} S.,  {Molendi} S.,  {Nagai} D.,  {Lau}
  E.~T.,  {Roncarelli} M.,  {Rossetti} M.,  {Snowden} S.~L.,    {Gastaldello}
  F.,  2012, \aap, 541, A57

\bibitem[\protect\citeauthoryear{{Ellison}, {Berezhko} \& {Baring}}{{Ellison}
  et~al.}{2000}]{ellison00}
{Ellison} D.~C.,  {Berezhko} E.~G.,    {Baring} M.~G.,  2000, \apj, 540, 292

\bibitem[\protect\citeauthoryear{{Federrath}, {Chabrier}, {Schober},
  {Banerjee}, {Klessen} \& {Schleicher}}{{Federrath}
  et~al.}{2011}]{federrath11}
{Federrath} C.,  {Chabrier} G.,  {Schober} J.,  {Banerjee} R.,  {Klessen}
  R.~S.,    {Schleicher} D.~R.~G.,  2011, Physical Review Letters, 107, 114504

\bibitem[\protect\citeauthoryear{{Finoguenov}, {Sarazin}, {Nakazawa}, {Wik} \&
  {Clarke}}{{Finoguenov} et~al.}{2010}]{fsn10}
{Finoguenov} A.,  {Sarazin} C.~L.,  {Nakazawa} K.,  {Wik} D.~R.,    {Clarke}
  T.~E.,  2010, \apj, 715, 1143

\bibitem[\protect\citeauthoryear{{Gaensler}, {Beck}, {Feretti} \&
  {Reich}}{{Gaensler} et~al.}{2008}]{gaensler08}
{Gaensler} B.~M.,  {Beck} R.,  {Feretti} L.,    {Reich} W.,  2008, in {Wada}
  K.,  {Combes} F.,  eds, Mapping the Galaxy and Nearby Galaxies {Revealing
  Cosmic Magnetism with the Square Kilometre Array}.
p.~323

\bibitem[\protect\citeauthoryear{{Giacalone} \& {Jokipii}}{{Giacalone} \&
  {Jokipii}}{2006}]{giacalone06}
{Giacalone} J.,  {Jokipii} J.~R.,  2006, Journal of Physics Conference Series,
  47, 160

\bibitem[\protect\citeauthoryear{{Giacintucci}, {Venturi}, {Macario},
  {Dallacasa}, {Brunetti}, {Markevitch}, {Cassano}, {Bardelli} \&
  {Athreya}}{{Giacintucci} et~al.}{2008}]{gvm08}
{Giacintucci} S.,  {Venturi} T.,  {Macario} G.,  {Dallacasa} D.,  {Brunetti}
  G.,  {Markevitch} M.,  {Cassano} R.,  {Bardelli} S.,    {Athreya} R.,  2008,
  \aap, 486, 347

\bibitem[\protect\citeauthoryear{{Hoeft} \& {Br{\"u}ggen}}{{Hoeft} \&
  {Br{\"u}ggen}}{2007}]{hoeft07}
{Hoeft} M.,  {Br{\"u}ggen} M.,  2007, \mnras, 375, 77

\bibitem[\protect\citeauthoryear{{Iapichino} \& {Br{\"u}ggen}}{{Iapichino} \&
  {Br{\"u}ggen}}{2012}]{iapichino12}
{Iapichino} L.,  {Br{\"u}ggen} M.,  2012, \mnras, 423, 2781

\bibitem[\protect\citeauthoryear{{Krymskii}}{{Krymskii}}{1977}]{1977DoSSR.234R%
1306K}
{Krymskii} G.~F.,  1977, Akademiia Nauk SSSR Doklady, 234, 1306

\bibitem[\protect\citeauthoryear{{Lee}, {Deane} \& {Federrath}}{{Lee}
  et~al.}{2009}]{lee09}
{Lee} D.,  {Deane} A.~E.,    {Federrath} C.,  2009, in {Pogorelov} N.~V.,
  {Audit} E.,  {Colella} P.,   {Zank} G.~P.,  eds, Numerical Modeling of Space
  Plasma Flows: ASTRONUM-2008 Vol.~406 of Astronomical Society of the Pacific
  Conference Series, {A New Multidimensional Unsplit MHD Solver in FLASH3}.
p.~243

\bibitem[\protect\citeauthoryear{{Macario}, {Markevitch}, {Giacintucci},
  {Brunetti}, {Venturi} \& {Murray}}{{Macario} et~al.}{2011}]{mmg11}
{Macario} G.,  {Markevitch} M.,  {Giacintucci} S.,  {Brunetti} G.,  {Venturi}
  T.,    {Murray} S.~S.,  2011, \apj, 728, 82

\bibitem[\protect\citeauthoryear{{Markevitch}, {Govoni}, {Brunetti} \&
  {Jerius}}{{Markevitch} et~al.}{2005}]{markevitch05}
{Markevitch} M.,  {Govoni} F.,  {Brunetti} G.,    {Jerius} D.,  2005, \apj,
  627, 733

\bibitem[\protect\citeauthoryear{{Nagai} \& {Lau}}{{Nagai} \&
  {Lau}}{2011}]{nagai11}
{Nagai} D.,  {Lau} E.,  2011, ArXiv e-prints

\bibitem[\protect\citeauthoryear{{Niemiec}, {Pohl}, {Stroman} \&
  {Nishikawa}}{{Niemiec} et~al.}{2008}]{niemiec08}
{Niemiec} J.,  {Pohl} M.,  {Stroman} T.,    {Nishikawa} K.-I.,  2008, \apj,
  684, 1174

\bibitem[\protect\citeauthoryear{{Ogrean}, {Br{\"u}ggen}, {van Weeren},
  {R{\"o}ttgering}, {Croston} \& {Hoeft}}{{Ogrean} et~al.}{2013}]{ogrean13}
{Ogrean} G.~A.,  {Br{\"u}ggen} M.,  {van Weeren} R.~J.,  {R{\"o}ttgering} H.,
  {Croston} J.~H.,    {Hoeft} M.,  2013, ArXiv e-prints

\bibitem[\protect\citeauthoryear{{Qin}, {Matthaeus} \& {Bieber}}{{Qin}
  et~al.}{2002}]{qin02}
{Qin} G.,  {Matthaeus} W.~H.,    {Bieber} J.~W.,  2002, \apjl, 578, L117

\bibitem[\protect\citeauthoryear{{Rybicki} \& {Lightman}}{{Rybicki} \&
  {Lightman}}{1986}]{rybicki86}
{Rybicki} G.~B.,  {Lightman} A.~P.,  1986, {Radiative Processes in
  Astrophysics}

\bibitem[\protect\citeauthoryear{{Schlickeiser}, {Sievers} \&
  {Thiemann}}{{Schlickeiser} et~al.}{1987}]{schlickeiser87}
{Schlickeiser} R.,  {Sievers} A.,    {Thiemann} H.,  1987, \aap, 182, 21

\bibitem[\protect\citeauthoryear{{Shin}, {Stone} \& {Snyder}}{{Shin}
  et~al.}{2008}]{shin08}
{Shin} M.-S.,  {Stone} J.~M.,    {Snyder} G.~F.,  2008, \apj, 680, 336

\bibitem[\protect\citeauthoryear{{Simionescu}, {Allen}, {Mantz}, {Werner},
  {Takei}, {Morris}, {Fabian}, {Sanders}, {Nulsen}, {George} \&
  {Taylor}}{{Simionescu} et~al.}{2011}]{simionescu11}
{Simionescu} A.,  {Allen} S.~W.,  {Mantz} A.,  {Werner} N.,  {Takei} Y.,
  {Morris} R.~G.,  {Fabian} A.~C.,  {Sanders} J.~S.,  {Nulsen} P.~E.~J.,
  {George} M.~R.,    {Taylor} G.~B.,  2011, ArXiv e-prints

\bibitem[\protect\citeauthoryear{{Subramanian}, {Shukurov} \&
  {Haugen}}{{Subramanian} et~al.}{2006}]{subramanian06}
{Subramanian} K.,  {Shukurov} A.,    {Haugen} N.~E.~L.,  2006, \mnras, 366,
  1437

\bibitem[\protect\citeauthoryear{{van Weeren}, {Br{\"u}ggen}, {R{\"o}ttgering}
  \& {Hoeft}}{{van Weeren} et~al.}{2011}]{vanweeren11}
{van Weeren} R.~J.,  {Br{\"u}ggen} M.,  {R{\"o}ttgering} H.~J.~A.,    {Hoeft}
  M.,  2011, \mnras, 418, 230

\bibitem[\protect\citeauthoryear{{van Weeren}, {R{\"o}ttgering}, {Br{\"u}ggen}
  \& {Hoeft}}{{van Weeren} et~al.}{2010}]{vanweeren10}
{van Weeren} R.~J.,  {R{\"o}ttgering} H.~J.~A.,  {Br{\"u}ggen} M.,    {Hoeft}
  M.,  2010, Science, 330, 347

\bibitem[\protect\citeauthoryear{{Vazza}, {Eckert}, {Simionescu}, {Br{\"u}ggen}
  \& {Ettori}}{{Vazza} et~al.}{2013}]{vazza13}
{Vazza} F.,  {Eckert} D.,  {Simionescu} A.,  {Br{\"u}ggen} M.,    {Ettori} S.,
  2013, \mnras, 429, 799

\bibitem[\protect\citeauthoryear{{Vink} \& {Laming}}{{Vink} \&
  {Laming}}{2003}]{vink03}
{Vink} J.,  {Laming} J.~M.,  2003, \apj, 584, 758

\bibitem[\protect\citeauthoryear{{V{\"o}lk}, {Berezhko} \&
  {Ksenofontov}}{{V{\"o}lk} et~al.}{2003}]{voelk03}
{V{\"o}lk} H.~J.,  {Berezhko} E.~G.,    {Ksenofontov} L.~T.,  2003, \aap, 409,
  563

\bibitem[\protect\citeauthoryear{{Zank}, {Li}, {Florinski}, {Hu}, {Lario} \&
  {Smith}}{{Zank} et~al.}{2006}]{zank06}
{Zank} G.~P.,  {Li} G.,  {Florinski} V.,  {Hu} Q.,  {Lario} D.,    {Smith}
  C.~W.,  2006, Journal of Geophysical Research (Space Physics), 111, 6108

\end{thebibliography}

\end{document}